\documentstyle[preprint,tighten,aps,floats,epsf,psfig,amssymb]{revtex}

\begin{document}

\newcommand{\be}{\begin{equation}}
\newcommand{\ee}{\end{equation}}
\newcommand{\bea}{\begin{eqnarray}}
\newcommand{\eea}{\end{eqnarray}}
\newcommand{\<}{\langle}
\renewcommand{\>}{\rangle}

\def\spose#1{\hbox to 0pt{#1\hss}}
\def\ltapprox{\mathrel{\spose{\lower 3pt\hbox{$\mathchar"218$}}
 \raise 2.0pt\hbox{$\mathchar"13C$}}}
\def\gtapprox{\mathrel{\spose{\lower 3pt\hbox{$\mathchar"218$}}
 \raise 2.0pt\hbox{$\mathchar"13E$}}}
\def\bsigma{\mbox{\protect\boldmath $\sigma$}}
\def\btau{\mbox{\protect\boldmath $\tau$}}
\def\bphi{\mbox{\protect\boldmath $\phi$}}
\def\bz{\mbox{\protect\boldmath $z$}}
\def\bw{\mbox{\protect\boldmath $w$}}
\def\hatp{\hat p}
\def\hatl{\hat l}
\def\smfrac#1#2{{\textstyle\frac{#1}{#2}}}
\def\case#1#2{{\textstyle\frac{#1}{#2}}}
\def\msbar{ {\overline{\hbox{\scriptsize MS}}} }
\def\normalmsbar{ {\overline{\hbox{\normalsize MS}}} }

\newcommand{\R}{\hbox{{\rm I}\kern-.2em\hbox{\rm R}}}
\newcommand{\reff}[1]{(\ref{#1})}

\title{Crossover phenomena in spin models with medium-range
       interactions and  self-avoiding walks with 
       medium-range jumps}
\author{
  \\
  {\small Sergio Caracciolo,${}^a$ Maria Serena Causo,${}^b$ 
          Andrea Pelissetto,${}^c$} \\
  {\small  Paolo Rossi,${}^d$ Ettore Vicari${}^d$} \\
  {\small\it ${}^a$ Scuola Normale Superiore, INFN and INFM, 
              I-56100 Pisa , ITALIA}  \\
  {\small\it ${}^b$ Scuola Normale Superiore and INFM,
              I-56100 Pisa , ITALIA} \\
  {\small\it ${}^c$ Dipartimento di Fisica, Universit\`a di Roma La Sapienza} 
         \\
  {\small\it and 
              INFN, Sezione di Roma,
              I-00185 Roma, ITALIA} \\
  {\small\it ${}^d$ Dipartimento di Fisica, Universit\`a di Pisa and 
              INFN, Sezione di Pisa,
              I-56100 Pisa, ITALIA} \\
}

\maketitle

\begin{abstract}
We study crossover phenomena in a model of self-avoiding walks with
medium-range jumps, that corresponds to the limit 
$N\to 0$ of an $N$-vector spin system with medium-range interactions.

In particular, we consider the critical crossover limit that
interpolates between the Gaussian and the Wilson-Fisher fixed point.
The corresponding crossover functions  are computed using 
field-theoretical methods and an appropriate mean-field expansion. 
 
The critical crossover limit is accurately studied by numerical Monte Carlo
simulations, which are much more efficient for walk models 
than for spin systems. 
Monte Carlo data are  compared with the field-theoretical predictions
concerning the critical crossover functions, finding a good
agreement. We also verify the predictions for the scaling behavior
of the leading nonuniversal corrections.

We determine phenomenological parametrizations that are exact
in the critical crossover limit, have the correct scaling behavior
for the leading correction, and describe the nonuniversal
crossover behavior of our data for any finite range.
\end{abstract}



\newpage

\section{Introduction}

The universality of critical phase transitions is related to the 
presence of a diverging correlation length $\xi$. When $\xi$ 
is much larger than any microscopic scale characterizing the system,
one observes a scaling behavior that is universal, i.e. independent 
of the microscopic details.  However, in experimental situations
the correlation length may not be so large and, on the contrary, it may 
be comparable to some 
other scale intrinsic to the system. In this case, one does not 
observe the expected critical behavior, but rather a crossover. 
Here, we will be interested in the crossover between the standard
Wilson-Fisher behavior (near the critical point) and the mean-field
behavior (far from the critical point)
that is observed by varying the 
temperature in systems belonging to the $N$-vector 
universality class, i.e. magnets, fluids, multicomponent fluid mixtures, ... 
(the critical behavior of these systems is reviewed, e.g., in 
Refs. \cite{Cargese,review}). 
Such a crossover is characterized by the 
Ginzburg number $G$ \cite{Ginzburg_60} that measures the relevance 
of the magnetization (or density) space fluctuations that determine 
the departure from the Landau mean-field behavior. If 
$t\equiv (\beta_c-\beta)/\beta_c$ is the reduced temperature, 
for $|t|>G$ the system shows an approximate mean-field 
behavior, while for $|t|<G$ one observes the standard 
Wilson-Fisher criticality. The crossover behavior is nonuniversal
since it depends on the specific details of the system under investigation, 
and is usually described in terms of phenomenological models
(see, e.g., Refs. 
\cite{Chen-etal_90,Chen-etal_90b,KKP-91,Anisimov-etal_92,Belyakov-Kiselev_92,%
JTS-93,KWAS-99,AS-review,KE-00,BLMWB-00,LM-00,LK-01,AAS-01} 
and references therein for a 
discussion of phenomenological models for fluids, binary mixtures, 
and polymers). 
However, in a specific limit---we call it critical crossover limit---one can 
define universal quantities that 
do not depend on the microscopic details. 

In this paper, we will consider spin models with medium-range interactions.
For instance, we may consider the Hamiltonian 
\be
{H} = - \sum_x \sum_{y:|y-x|\le R} \sigma_x\cdot \sigma_{y},
\label{eq1.1}
\ee
where $\sigma_x$ is an $N$-component vector satisfying $\sigma_x\cdot\sigma_x=1$. 
The crossover behavior of these systems has been extensively studied
numerically 
\cite{M-B-93,L-B-B-pre,L-B-B-prl,Luijten-Binder_98,Luijten,OPF-00,BLMWB-00}.
By means of scaling arguments, it was shown \cite{M-B-93,L-B-B-pre,L-B-B-prl}
that the Ginzburg number $G$ is proportional to $R^{-2d/(4-d)}$ 
in $d$ dimensions. Thus,
for $|t| \gg R^{-2d/(4-d)}$
such systems show an approximate mean-field behavior, while for $|t|\ll
R^{-2d/(4-d)}$
one observes the standard Wilson-Fisher criticality.

Such a crossover can be described by using
effective exponents. For instance, one can define an effective susceptibility
exponent $\gamma_{\rm eff}(t)$ (often called Kouvel-Fisher exponent 
\cite{KF-64}) by
\be 
\gamma_{\rm eff}(t,R) = - {t\over \chi_R(t)} {d\chi_R(t)\over dt},
\ee
where $\chi_R(t)$ is the susceptibility. By varying the temperature, 
the exponent $\gamma_{\rm eff}(t,R)$ varies between 1, the mean-field value, 
and $\gamma$, the Wilson-Fisher value (actually, the full crossover behavior 
can be observed only for $R$ large enough, in lattice models for $R\gtapprox 3$). 
At $R$ fixed the crossover 
behavior is not universal, and therefore the function $\gamma_{\rm eff}(t,R)$
at fixed $R$ cannot be predicted without an explicit reference to the 
microscopic details of the system. In this case, if one wishes to 
obtain interpolations that provide reasonably precise 
approximations, one must resort to phenomenological models, such as those 
presented in Refs. 
\cite{Chen-etal_90,Chen-etal_90b,KKP-91,Anisimov-etal_92,Belyakov-Kiselev_92,%
JTS-93,KWAS-99,AS-review,KE-00,BLMWB-00,LM-00,LK-01}.
However, there is a particular limit---the critical
crossover limit---in which the effective exponents 
become universal. If we consider the limit
$t\to 0$, $R\to \infty$ with $\tilde{t}\equiv t/G \sim t R^{2d/(4-d)}$ fixed, 
then $\gamma_{\rm eff}(t,R)$ converges to a critical crossover exponent 
$\gamma_{\rm eff}(\tilde{t})$ that is universal
apart from a trivial rescaling of $\tilde{t}$. In practice, at least 
in lattice models such as \reff{eq1.1}, the critical crossover 
functions provide a good description of the crossover as soon as the 
interactions extend over a few (two or three) lattice spacings.

The universality of the critical crossover functions can be shown explicitly 
in the large-$N$ limit \cite{PRV-1,PRV-2} and for any value of $N$ by 
performing an expansion around mean field \cite{PRV-2}. Moreover, as 
shown in Ref. \cite{PRV-2}, they are given {\em exactly} by the 
field-theoretical crossover curves computed within the 
$\phi^4$ framework in Refs. 
\cite{Bagnuls-Bervillier_84,Bagnuls-Bervillier_85,BBMN-87,Schloms-Dohm_89,%
K-S-D}. 

In this paper we wish to check with a high-precision simulation the 
field-theoretical predictions of Ref. \cite{PRV-2}. Large-scale 
Monte Carlo results have already been obtained for the Ising model, both
in two and in three dimensions
\cite{L-B-B-pre,L-B-B-prl,Luijten-Binder_98,Luijten}. However, because of the 
difficulty of keeping under control the finite-size effects, only small 
ranges $R$ were simulated near the Wilson-Fisher point. Although the 
general trend of the data was consistent with the analytic field-theoretical
predictions, still in the Wilson-Fisher region a high-precision 
numerical check could not be done. The considered values of $R$  
were too small 
and there were significant discrepancies between numerical data 
and theoretical predictions.

Here, we address the problem for a spin model in the limit 
$N\to 0$, which can be described in terms of self-avoiding walks (SAWs)
\cite{deGennes_72,Daoud-etal_75,Emery_75,ACF-83,deGennes_book,%
desCloizeaux-Jannink_book}. The advantage of such a system is that we can now 
work directly in the infinite-volume limit without 
finite-size effects and thus we can investigate systems with 
much larger values of the 
correlation length (in this paper we reach $\xi\approx 500$ 
for systems in which the interaction extends up to 12 lattice spacings). 
In the limit $N\to 0$,
the model \reff{eq1.1} is mapped into a model of SAWs with 
medium-range jumps, i.e. of SAWs such that the length of each link
is less than or equal to $R$. As usual in walk simulations, we work 
in a monodisperse ensemble, i.e. with walks of fixed length $n$. The 
length $n$ replaces here the reduced temperature $t$.
Medium-range SAWs show a crossover behavior depending on 
$n G\sim n R^{-2d/(4-d)}$. For $n \ll R^{2d/(4-d)}$ the SAW
behaves as an ordinary random walk (mean-field behavior), while in the 
opposite regime the self-repulsion becomes important and one observes the 
standard critical behavior. As we already stressed, for fixed values of $R$, 
such a behavior is not universal and can only be described 
phenomenologically. There is however a universal limit, the 
{\em critical} crossover limit: If we take the limit 
$n\to\infty$, $R\to\infty$ keeping 
$\widetilde{n} = n R^{-2d/(4-d)}$ fixed, the crossover functions become
universal, and can again be computed by using field-theory methods.

The paper is organized as follows. 

In Sec. \ref{sec2} 
we introduce the model and the basic observables we consider.

In Sec. \ref{sec3} we discuss the critical crossover limit. 
In Sec. \ref{sec3.1} we define the limit in spin models 
\cite{L-B-B-pre,L-B-B-prl,Luijten-Binder_98} and review the 
results of Ref. \cite{PRV-2}. In Sec. \ref{sec3.2} we define 
the crossover limit for walk models and derive some general results 
for the universal crossover functions. In particular, we show 
that they are {\em exactly} related to the crossover 
functions computed in field theory 
\cite{Muthukumar-Nickel_84,Muthukumar-Nickel_87,dCCJ-85}.
Moreover, by using the 
results of Ref. \cite{PRV-2}, we show that by an appropriate definition
of the range $R$ the leading corrections to the universal crossover 
functions scale
as $R^{-d}$, $d$ being the dimension, as $R\to \infty$. 

In Sec. \ref{sec4}
we derive the expressions of the crossover functions from field theory 
generalizing the results of Ref. \cite{Muthukumar-Nickel_87}. 
Details are reported in the Appendix, where we also compute 
the first coefficients of the asymptotic expansion of the crossover 
functions near the Wilson-Fisher point by using the fixed-dimension 
expansion in the zero-momentum scheme \cite{Parisi_Cargese} and in the 
dimensional regularization scheme without $\epsilon$-expansion 
\cite{Dohm,Schloms-Dohm_89,K-S-D}. 

In Sec. \ref{sec5} 
we briefly describe the numerical algorithms we use.

In Sec. \ref{sec6} we perform a
detailed comparison of the numerical results in three dimensions
with the field-theory predictions. 
We find a very good agreement, the deviations being small already when
the interaction extends over three lattice spacings. Particular care 
has been devoted to the behavior of the leading corrections. We show
that they scale as $R^{-d}$ as predicted in 
Ref. \cite{PRV-2}. 

Finally, in Sec. \ref{sec7} we report our conclusions
and discuss some further applications of these results. In particular, we give 
phenomenological expressions that are able
to describe the crossover curves even outside the critical limit, for all 
ranges $R$ we have considered.

Preliminary results appeared in Ref. \cite{CCPRV-99}.

\section{The model} \label{sec2}

In this paper we consider SAWs
with medium-range jumps. To be specific, let us consider 
a hypercubic lattice in $d$ dimensions. Given an integer number 
$\rho$, let us define a lattice domain $D_\rho(x)$. 
If $x$ is a lattice point, $D_\rho(x)$ is the 
set of lattice points defined by 
\be
D_\rho(x) =\, \left\{ y: \sum_{i=1}^d |x_i - y_i| \le \rho\right\}.
\label{Dshape}
\ee
We indicate with $V_\rho$
the number of points belonging to $D_\rho(x)$ and with $R$
the mean square size of $D_\rho(x)$. Explicitly, we define
\begin{eqnarray}
V_\rho &\equiv& \sum_{y\in D_\rho(0)} 1\; , 
\label{defVrho} \\
R^2 &\equiv& {1\over 2 d V_\rho} \sum_{y\in D_\rho(0)} y^2 \; .
\label{defR2}
\end{eqnarray}
In three dimensions
\begin{eqnarray}
V_\rho &=& {1\over 3} (2 \rho + 1) (2 \rho^2 + 2 \rho + 3), \\
R^2 &=& 
    {\rho(\rho+1)\over10} {\rho^2 + \rho + 3\over 2 \rho^2 + 2 \rho + 3}.
\end{eqnarray}
For $\rho\to\infty$, $V_\rho \approx {4\over3} \rho^3$ and 
$R^2\approx {1\over20} \rho^2$.
In the following we will often characterize the size of the jumps by 
$R$ instead of $\rho$ and thus we will write 
$D_R$, $V_R$, $\ldots$, instead of $D_\rho$, $V_\rho$, $\ldots$

Let us now define our model.
We define an $n$-step $R$-SAW as a sequence of lattice 
points $\{\omega_0,\ldots,\omega_n\}$ with 
$\omega_0 = (0,\ldots,0)$ and $\omega_{j+1} \in D_R(\omega_j)$, such that
$\omega_i\not=\omega_j$ for all $i\not= j$. All walks are weighted 
equally. For $\rho=1$ the model corresponds to a standard SAW with 
nearest-neighbor jumps.

We will consider the following observables: 
If $c_{n,R}(x)$ is the number of $n$-step $R$-SAWs going from $0$ to $x$,
we indicate with $c_{n,R}$ 
the total number of $n$-step walks and with $E^2_{n,R}$ 
the mean square end-to-end distance. They are defined as follows:
\begin{eqnarray}
c_{n,R} &\equiv& \sum_x c_{n,R}(x), \\
E^2_{n,R} &\equiv& {1\over c_{n,R}} \sum_x x^2 c_{n,R}(x).
\end{eqnarray}
This model of walks is related 
to a lattice $N$-vector model with 
medium-range interactions in the limit $N\to 0$. Indeed, consider 
the Hamiltonian
\be
H_R({\sigma}) = - {N\over 2} \sum_x \sum_{y\in D_R(x)}
    {\sigma}_x \cdot {\sigma}_y,
\label{HNvector}
\ee
where ${\sigma}_x$ is an $N$-dimensional vector satisfying 
${\sigma}_x \cdot {\sigma}_x= 1$, and define as usual
\begin{eqnarray}
Z_{R}(\beta) &\equiv& \sum_{\{\sigma\}} e^{- \beta H_R({\sigma})} ,\\
G_R(x;\beta) &\equiv& \< {\sigma}_0 \cdot {\sigma}_x \>_R =\,
  {1\over Z_{R}(\beta)} 
  \sum_{\{\sigma\}} {\sigma}_0 \cdot {\sigma}_x \ 
     e^{- \beta H_R({\sigma})}.
\end{eqnarray}
The susceptibility and the (second-moment) 
correlation length are then defined as 
\begin{eqnarray}
\chi_R(\beta) &\equiv& \sum_x G_R(x;\beta),   \\
\xi^2_R(\beta) &\equiv& {1\over 2 d \chi_R(\beta)} \sum_x x^2 G_R(x;\beta).
\end{eqnarray}
A standard procedure 
\cite{deGennes_72,Daoud-etal_75,Emery_75,ACF-83,deGennes_book,%
desCloizeaux-Jannink_book} 
allows to prove that 
\begin{eqnarray}
\lim_{N\to 0} \chi_R(\beta) &= &
     \sum_{n=0}^\infty \beta^n c_{n,R} \; ,
\label{chiRNeq0} \\
\lim_{N\to 0}  \xi^2_R(\beta) \chi_R(\beta) &= & {1\over 2d}
    \sum_{n=0}^\infty \beta^n c_{n,R} E^2_{n,R}.
\label{xiRNeq0}
\end{eqnarray}
This equivalence will allow us to use the results available for 
the Hamiltonian \reff{HNvector} that are discussed in detail 
in Ref. \cite{PRV-2}.

\section{Critical crossover limit} \label{sec3}

In this Section we derive some general results for the 
critical crossover limit of medium-range SAWs. They will 
be obtained by extending to walk models the results of Ref. \cite{PRV-2}.

\subsection{The variable-length $\beta$-ensemble} \label{sec3.1}

Let us consider the Hamiltonian \reff{HNvector}, which, for 
$R$ fixed, defines a generalized $N$-vector
model with short-range interactions. 
For each value of $R$ there is a critical point\footnote{For the 
general $N$-vector model in two dimensions, a 
critical point exists only for $N\le 2$. Theories with $N\ge 3$ are 
asymptotically free and become critical only in the limit $\beta\to \infty$.}
$\beta_{c,R}$; for $\beta\to \beta_{c,R}$ the susceptibility and the 
correlation length have the standard behavior
\begin{eqnarray}
\chi_R(\beta) &\approx& A_\chi(R) t^{-\gamma} (1 + B_\chi(R) t^\Delta + 
       \cdots) ,
\label{chiRfisso} \\
\xi_R^2(\beta) &\approx& A_\xi(R) t^{-2\nu} (1 + B_\xi (R) t^\Delta + 
       \cdots),
\label{xiRfisso} 
\end{eqnarray}
where $t \equiv (\beta_{c,R} - \beta)/\beta_{c,R}$ and we have neglected 
additional subleading corrections.
The exponents $\gamma$, $\nu$, and $\Delta$ do not depend on $R$. 
In two dimensions, for $N=0$, $\gamma$ and $\nu$ are known exactly
\cite{Nienhuis_82_84}
\be
\nu    =\, {3\over4}, \qquad\qquad
\gamma =\, {43\over 32},
\ee
while $\Delta$ is still the object of an intense debate
\cite{Guttmann-Enting_88,Conway-Guttmann_96,Saleur_87,Conway-Guttmann_93,%
Guim-etal_97,Caracciolo-etal_98}.
In three dimensions, for $N=0$, the best estimates of the exponents 
have been obtained in Monte Carlo simulations:
\begin{eqnarray}
\nu &=& \cases{0.5877 \pm 0.0006 & \hskip 1truecm Ref. \cite{Li-etal_95},   \cr
               0.58758 \pm 0.00007  & 
               \hskip 1truecm Ref. \cite{Belohorec-Nickel_97},  } \nonumber \\ 
\gamma &=& 1.1575 \pm 0.0006 
      \hbox{\hskip 2.15truecm Ref. \cite{Caracciolo-etal_97}, } 
\label{esponenticritici}\\
\Delta &=& 0.515 {}^{+0.017}_{-0.007}
      \hbox{\hskip 3.1truecm Ref. \cite{Belohorec-Nickel_97}.  } \nonumber 
\end{eqnarray}
Less precise Monte Carlo results can be found in Refs.  
\cite{Grassberger-etal_97,Pedersen-etal_96,Eizenberg-Klafter_96,BGZ-00}
and references therein.
Similar, although less precise, results are obtained by using 
field-theory methods and from the analysis of enumeration series
(for a list of results, see Refs.
\cite{Muthukumar-Nickel_87,Murray-Nickel_91,%
Guida-ZinnJustin_98,BTW-99,JK-01,Butera-Comi_97,MJHMJG-00} and
references therein). 

On the other hand, the 
amplitudes are nonuniversal and depend on $R$.
For $R\to\infty$, they behave as \cite{L-B-B-pre,L-B-B-prl}
\begin{eqnarray} 
A_\chi(R) &\approx &   A_\chi^\infty R^{2 d(1-\gamma)/(4 - d)} ,\qquad
A_\xi(R) \approx A_\xi^\infty   R^{4 (2 - d\nu)/(4-d)},  \nonumber \\ 
B_\chi(R) &\approx &   B_\chi^\infty R^{2 d \Delta/(4 - d)} ,\qquad \qquad
B_\xi(R) \approx  B_\xi^\infty  R^{2 d \Delta /(4-d)} .
\label{eq3.4}
\end{eqnarray}
Corrections to these asymptotic behaviors vanish as \cite{PRV-2} $R^{-d}$.

The critical point $\beta_{c,R}$ also depends on $R$. 
The expansion of $\beta_{c,R}$ for $R\to \infty$ was derived in 
Ref. \cite{PRV-2} in two and three dimensions. Explicitly, for $N=0$
and $d=3$, we have
\be
\beta_{c,R} = {1\over V_R}\left(1 + \overline{I}_R - 
        {3\over 32\pi^2 R^6} \log R^2 + {\tau_1\over R^6} + 
        {\tau_2\over R^8}
          + O(R^{-9}\log R^2)\right),
\label{betacR-3d}
\ee
where $\tau_1$ and $\tau_2$ are constants and $\overline{I}_R$ is a 
function of $R$. 
The nonperturbative constants $\tau_1$ and $\tau_2$ depend on the domain. 
The expression of $\tau_1$ for a 
generic domain is reported in Ref. \cite{PRV-2}. For the domain \reff{Dshape},
$\tau_1 \approx -0.00060(11)$. The constant $\tau_2$ will be computed 
numerically in Sec. \ref{sec6.1}.  
The function $\overline{I}_R$ is defined by
\be
\overline{I}_R \equiv \int {d^3k\over (2\pi)^3} {1 - \Pi_R(k)\over \Pi_R(k)},
\ee
where 
\be
\Pi_R(k) \equiv  1 - {1\over V_R} \sum_{x\in D_R(0)} e^{ik\cdot x}.
\ee
For $R\to \infty$, $\overline{I}_R\approx \sigma R^{-3} + O(R^{-5})$. For the 
domain considered in this paper $\sigma\approx 0.04336529$. Explicit values of 
$\overline{I}_R$ are reported in Table \ref{IbarR}. 
\begin{table}
\caption{\label{IbarR}
Estimates of $R^3\overline{I}_{R}$ for several values of $\rho$ 
for the domain \protect\reff{Dshape}. From Ref. \protect\cite{PRV-2}.
 }
\begin{tabular}{cccc}
$\rho$ & $R^3\overline{I}_{R}\vphantom{\displaystyle{1\over2}}$ & 
$\rho$ & $R^3\overline{I}_{R}$ \\
\hline
3  & 0.043960387 & 10 & 0.043486698  \\
4  & 0.043921767 & 12 & 0.043451767  \\
5  & 0.043713672 & 14 & 0.043429899  \\
6  & 0.043664053 & 16 & 0.043415345  \\
7  & 0.043574469 & 18 & 0.043405187  \\
8  & 0.043547206 & 20 & 0.043397824  \\
\end{tabular}
\end{table} 

Let us now define the critical crossover limit. 
In this case we consider the limit 
$R\to\infty$, $t\to0$, with 
$\widetilde{t} \equiv R^{2d/(4-d)}t$ fixed.
It is possible to show that
\begin{eqnarray}
\widetilde{\chi}_R \equiv 
         R^{-2d/(4-d)} \chi_R(\beta) &\to& f_\chi(\widetilde{t}) ,
\label{fchi} \\
\widetilde{\xi}^2_R \equiv 
R^{-8/(4-d)} \xi^2_R(\beta) &\to& f_\xi(\widetilde{t}) ,
\label{fxi}
\end{eqnarray}
where the functions $f_\chi(\widetilde{t})$ and 
$f_\xi(\widetilde{t})$ are universal apart from an overall 
rescaling of $\widetilde{t}$ and a constant factor. 
Equations \reff{fchi} and \reff{fxi} were predicted in 
Refs. \cite{L-B-B-pre,L-B-B-prl} by means of a scaling argument 
and were proved to all orders in an expansion around 
$\widetilde{t}=\infty$ in Ref. \cite{PRV-2}.
 
The crossover functions have a well-defined behavior in the limiting
cases $\widetilde{t}\to 0$ and $\widetilde{t}\to \infty$.
For $\widetilde{t}\to 0$, Eqs. \reff{chiRfisso} and \reff{xiRfisso} imply
\begin{eqnarray}
f_\chi(\widetilde{t}) &\approx & A_\chi^\infty \widetilde{t}^{-\gamma}
    (1 + B_\chi^\infty \widetilde{t}^{\Delta} + \ldots) , 
\label{fchi-Wilson-Fisher}\\
f_\xi(\widetilde{t}) &\approx & A_\xi^\infty \widetilde{t}^{-2 \nu}
    (1 + B_\xi^\infty \widetilde{t}^{\Delta} +\, \ldots) \; .
\label{fxi-Wilson-Fisher}
\end{eqnarray}
In the limit $\widetilde{t}\to \infty$, for generic values of $d$,
the crossover functions behave as 
\begin{eqnarray}
f_\chi(\widetilde{t}) &\approx& {a_\chi\over \widetilde{t}} 
  \left(1 + \alpha_\chi \widetilde{t}^{-\Delta_{\rm mf}} + 
   O(\widetilde{t}^{-2 \Delta_{\rm mf}}) \right), 
\label{fchimf}\\
f_\xi(\widetilde{t}) &\approx& {a_\xi\over \widetilde{t}} 
  \left(1 + \alpha_\xi \widetilde{t}^{-\Delta_{\rm mf}}
 + O(\widetilde{t}^{-2 \Delta_{\rm mf}}) \right),
\label{fximf}
\end{eqnarray}
where $\Delta_{\rm mf} = (4-d)/2$. It is important to notice that this 
expansion is corrected by logarithms whenever $d=4-2/k$, 
$k$ integer,
and therefore in the interesting cases $d=2,3$. For $d=3$ the neglected terms
in Eqs. \reff{fchimf} and \reff{fximf} are of order 
$O(\widetilde{t}^{-1} \log\widetilde{t})$ and not simply of order
$O(\widetilde{t}^{-1})$.
A detailed derivation of these expansions and of the expressions
(\ref{fchimf}) and \reff{fximf} is given in Ref. \cite{PRV-2} for a 
much more general model than the one considered here.
The constants $a_\chi$, $a_\xi$, $\alpha_\chi$, and $\alpha_\xi$ are given,
for $2<d<4$, by
\begin{eqnarray}
&& a_\chi = a_\xi = 1, \nonumber \\
&& \alpha_\chi = \alpha_\xi = - (4 \pi)^{-d/2}\Gamma(1 - d/2).
\label{constants-MF}
\end{eqnarray}
Additional terms can be computed exactly by using the field-theoretical 
results of Refs. \cite{Bagnuls-Bervillier_84,Bagnuls-Bervillier_85}, 
the perturbative series of Refs. \cite{Baker-etal_77_78,Murray-Nickel_91}, 
and the mean-field results of Ref. \cite{PRV-2}, see Sec. \ref{sec4}. 

It is also possible to compute the 
corrections to Eqs. \reff{fchi} and \reff{fxi}. On the basis 
of a two-loop calculation, Ref. \cite{PRV-2} conjectured that, if the 
range is expressed in terms of the variable $R$ defined in 
Eq. \reff{defR2},\footnote{
This behavior
can be observed only for quantities that are defined
using $R$ as scale. If we were considering, for instance,
$\chi_R(t) \rho^{-2 d/(4-d)}$ we would of course obtain the same universal
limiting curve, but now with corrections of order $1/\rho$.} 
then the corrections~\footnote{
 This behavior is correct for 
      $2<d<4$. In two dimensions, the leading correction cannot be 
      computed in a mean-field expansion and thus we do not 
      have a theoretical prediction. However, numerical and 
      theoretical arguments \cite{L-B-B-pre,L-B-B-prl,PRV-2} indicate that the 
      corrections should be of order $1/R^{-2}$ times 
      powers of $\log R$.}
scale as $R^{-d}$. Explicitly, in the 
critical crossover limit we expect
\begin{eqnarray}
\widetilde{\chi}_R \to f_\chi(\widetilde{t}) + {1\over R^d} 
                       h_\chi(\widetilde{t}) + \cdots 
\label{fchi-corrections}
\\
\widetilde{\xi}_R \to f_\xi(\widetilde{t}) + {1\over R^d}
                       h_\xi(\widetilde{t}) + \cdots
\label{fxi-corrections}
\end{eqnarray}
For $\widetilde{t}\to 0$ and $\widetilde{t}\to \infty$ the functions
$h_\chi(\widetilde{t})$ and $h_\xi(\widetilde{t})$ have an asymptotic
behavior that is analogous to that of the universal crossover 
functions $f_\chi(\widetilde{t})$ and $f_\xi(\widetilde{t})$. 
In Ref. \cite{PRV-2} the leading term for $\widetilde{t}\to \infty$ 
was computed, obtaining
\be
h_\chi(\widetilde{t}) \approx - {E_d\over\widetilde{t}}, \qquad
  h_\xi(\widetilde{t}) \approx - {E_d\over\widetilde{t}},
\label{hlargettilde}
\ee
where $E_d$ is a domain-dependent constant (see Ref. \cite{PRV-2} for 
its definition). For $d=3$ and for the 
domain \reff{Dshape}, we have $E_3 \approx 0.058545$.

\subsection{The fixed-length ensemble} \label{sec3.2}

Given the previous results it is now a completely standard 
procedure \cite{desCloizeaux-Jannink_book} 
to obtain the behavior of $c_{n,R}$ and $E^2_{n,R}$. 
For $n\to\infty$ at $R$ fixed, we obtain the standard behavior
\begin{eqnarray}
c_{n,R} &\approx& C_\chi(R)\,  \beta_{c,R}^{-n}\,  n^{\gamma - 1} 
     (1 + D_\chi(R) n^{-\Delta} + \cdots ), 
\label{cnRasym} \\
E^2_{n,R} &\approx& C_E(R)\, n^{2 \nu}  (1 + D_E(R) n^{-\Delta} + \cdots),
\end{eqnarray}
where
\begin{eqnarray}
C_\chi(R) &=& {A_\chi(R)\over \Gamma(\gamma)}, \nonumber\\
C_E(R)  &=& {2 d A_\xi(R) \Gamma(\gamma)\over \Gamma(\gamma + 2 \nu)}, 
 \nonumber \\
D_\chi(R) &=& {B_\chi(R) \Gamma(\gamma)\over \Gamma(\gamma - \Delta)}, 
 \nonumber \\
D_E(R) &=& {(B_\chi(R) + B_\xi(R))\Gamma(\gamma + 2\nu )\over 
           \Gamma(\gamma + 2 \nu - \Delta)} - 
          {B_\chi(R) \Gamma(\gamma)\over \Gamma(\gamma - \Delta)}.
\label{relazioni-ampiezze}
\end{eqnarray}
For $R\to\infty$, using Eq. \reff{eq3.4}, we obtain 
$C_\chi(R)\to C_\chi^\infty R^{2d(1-\gamma)/(4-d)}$,  where 
$C_\chi^\infty = A_\chi^\infty/\Gamma(\gamma)$, with corrections 
of relative order $R^{-d}$. Similar relations
hold for the other amplitudes.

The critical crossover limit is trivially defined by remembering that $n$ 
is the dual variable (in the sense of Laplace transforms) of 
$t$. Therefore, we should study the limit
$n\to\infty$, $R\to\infty$ with $\widetilde{n}\equiv n R^{-2d/(4-d)}$ fixed.
From Eqs. \reff{fchi} and \reff{fxi} we obtain that the following limits 
exist:
\begin{eqnarray}
&& \widetilde{c}_{n,R} \equiv \; 
c_{n,R} \beta_{c,R}^n \to g_c(\widetilde{n}), 
\label{defctilden}\\
&& \widetilde{E}^2_{n,R} \equiv \;
E^2_{n,R} R^{-8/(4-d)} \to g_E(\widetilde{n}), 
\label{defEtilden}
\end{eqnarray}
where the functions $g_c(\widetilde{n})$ and $g_E(\widetilde{n})$ are related 
by a Laplace transform to $f_\chi(\widetilde{t})$ and 
$f_\xi(\widetilde{t})$. Explicitly 
\begin{eqnarray}
f_\chi(t) &=& \int^\infty_0 du\,g_c(u) e^{-ut},  
\label{fchi-gc}  \\
f_\xi(t) f_\chi(t) &=& {1\over 2d} \int^\infty_0 du\, g_c(u) g_E(u) e^{-ut}.  
\label{fxi-gE}
\end{eqnarray}
Notice that, while
the knowledge of $\beta_{c,R}$ is not required for the definition 
of $g_E(\widetilde{n})$, the critical point is needed to 
compute $g_c(\widetilde{n})$.

The standard critical behavior is obtained for
$\widetilde{n}\to\infty$. In this limit we have
\begin{eqnarray}
g_c(\widetilde{n}) &\approx& C_\chi^\infty 
       \widetilde{n}^{\gamma-1}(1 + D_\chi^\infty \widetilde{n}^{-\Delta}
         + \cdots ),
\label{gc-WF}
\\
g_E(\widetilde{n}) &\approx& C_E^\infty 
       \widetilde{n}^{2 \nu}(1 + D_E^\infty \widetilde{n}^{-\Delta} 
         + \cdots ).
\label{gE-WF}
\end{eqnarray}
The mean-field limit corresponds to $\widetilde{n}\to 0$. 
Using Eqs. \reff{fchimf} and \reff{fximf}, we obtain
\begin{eqnarray}
g_c(\widetilde{n}) &\approx& 1 + \zeta_c \widetilde{n}^{\Delta_{\rm mf}} 
        + \cdots  
\label{gcmf}\\
g_E(\widetilde{n}) &\approx&  2 d\,  \widetilde{n} \;
    \left(1 + \zeta_E \widetilde{n}^{\Delta_{\rm mf}} + \cdots \right),
\label{gEmf}
\end{eqnarray}
with corrections of order $\widetilde{n}^{2 \Delta_{\rm mf}}$. 
In two and three dimensions additional logarithms appear. For $d=3$
the neglected corrections to $g_c(\widetilde{n})$ in 
Eq. \reff{gcmf} are of order $\widetilde{n}\log \widetilde{n}$. 
However, it can be shown by using the field-theoretical results 
of App. \ref{secA.3} that the logarithmic terms exponentiate and that one can 
write 
\be
g_c(\widetilde{n}) = e^{\zeta_{\rm log} \widetilde{n} \log \widetilde{n}}
          g_{c,\rm log}(\widetilde{n}),
\ee
where $\zeta_{\rm log}$ is a constant and $g_{c,\rm log}(\widetilde{n})$ is
a function with a regular expansion in powers of 
$\widetilde{n}^{\Delta_{\rm mf}}$ without logarithms. 
The behavior of $g_E(\widetilde{n})$ is simpler: It has a regular 
expansion in powers of $\widetilde{n}^{\Delta_{\rm mf}}$ in all 
dimensions without logarithms \cite{Muthukumar-Nickel_84}.
The constants $\zeta_c$ and $\zeta_E$ can be
easily related to $\alpha_\chi$ and $\alpha_\xi$ defined in 
Eqs. \reff{fchimf} and \reff{fximf}:
\begin{eqnarray}
\zeta_c &=& {\alpha_\chi\over \Gamma(1 + \Delta_{\rm mf})}, \\
\zeta_E &=& {\alpha_\chi + \alpha_\xi\over \Gamma(2 + \Delta_{\rm mf})} -
            {\alpha_\chi\over \Gamma(1 + \Delta_{\rm mf})}.
\end{eqnarray}
Using the explicit results \reff{constants-MF},
we obtain in three dimensions
\be
\zeta_c = \, {1\over 2 \pi^{3/2}}, \qquad\qquad
\zeta_E = \, {1\over 6 \pi^{3/2}}.
\ee
We wish now to compute the corrections to the universal crossover functions.
For $R\to\infty$, in the variable-length ensemble, 
the corrections are  $O(R^{-d})$, 
see Eqs. \reff{fchi-corrections} and
\reff{fxi-corrections}, for $2<d<4$. Thus, we expect that the universal 
crossover functions in the fixed-length ensemble have the same behavior. 
Therefore, we write
\begin{eqnarray}
\widetilde{c}_{n,R} \to g_c(\widetilde{n}) + {1\over R^d} k_c(\widetilde{n}),
\\
\widetilde{E}^2_{n,R} \to g_E(\widetilde{n}) + 
         {1\over R^d} k_E(\widetilde{n}).
\end{eqnarray}
It is easy to verify by using the Euler-Mac Laurin formula that 
\begin{eqnarray}
&& h_\chi(t) = \int_0^\infty du\, k_c(u) e^{-ut}, \\
&& f_\xi(t) h_\chi(t) + f_\chi(t) h_\xi(t) =\, 
    {1\over 2d}  \int_0^\infty du\, \left[g_E(u) k_c(u) + g_c(u) k_E(u)\right]
     e^{-ut}.
\end{eqnarray}
The asymptotic behavior of the correction functions $k_c(\widetilde{n})$
and $k_E(\widetilde{n})$ for $\widetilde{n}\to 0$ and $\widetilde{n}\to
\infty$ is analogous to that of $g_c(\widetilde{n})$
and $g_E(\widetilde{n})$. For $\widetilde{n}\to 0$, by using 
Eqs. \reff{hlargettilde}, \reff{fchimf}, and \reff{fximf}, we obtain
\be
k_c(0) = - E_d, \qquad\qquad
k_E(\widetilde{n}) = -2 d E_d \widetilde{n} + 
       O\left(\widetilde{n}^{1+\Delta_{\rm mf}}\right).
\label{corrections-k}
\ee

\section{Field-theory results in three dimensions} \label{sec4}

We wish now to compute the crossover functions by using 
field-theory methods. Consider the
continuum $\phi^4$ theory
\be
H = \int d^3 x\left[ {1\over2} (\partial_\mu \phi)^2 + 
                     {r\over2} \phi^2 + 
                     {u\over 4!} \phi^4 \right],
\ee
where $\phi$ is an $N$-dimensional vector---in our case $N=0$---, 
and introduce the Ginzburg
number $G\equiv u^{2/(4-d)}$ and $t\equiv r - r_c$ where 
$r_c$ is the critical value of $r$. Then, consider the limit
$u\to 0$, $t\to 0$, with $\widetilde{t}_{\rm SR}\equiv t/G = t u^{-2/(4-d)}$
fixed.  In this limit we have 
\bea
\widetilde{\chi}  &\equiv& \chi\, G \to\, F_\chi (\widetilde{t}_{\rm SR}), \\
\widetilde{\xi}^2 &\equiv& \xi^2\, G \to\, F_\xi (\widetilde{t}_{\rm SR}). 
\eea
The functions $F_\chi (\widetilde{t}_{\rm SR})$ and 
$F_\xi (\widetilde{t}_{\rm SR})$ can be computed by resumming appropriately the 
perturbative series. There are essentially two different perturbative series
one can consider: (a) the 
fixed-dimension expansion \cite{Parisi_Cargese,Bagnuls-Bervillier_84,%
Bagnuls-Bervillier_85}, which is at present the most precise
one since seven-loop series are available 
\cite{Baker-etal_77_78,Murray-Nickel_91}; 
(b) the so-called dimensional regularization without $\epsilon$-expansion
\cite{Dohm,Schloms-Dohm_89,K-S-D} that uses five-loop 
$\epsilon$-expansion results \cite{C-G-L-T,K-N-S-C-L}. In these two schemes 
the crossover functions are expressed in terms of various 
renormalization-group quantities. The explicit expressions are reported 
in App. \ref{secA.1} and \ref{secA.2}. 
For our purposes the relevant result is that 
the functions $F_\chi (\widetilde{t}_{\rm SR})$ and 
$F_\xi (\widetilde{t}_{\rm SR})$ are related by simple 
rescalings to the crossover functions we have defined before 
\cite{PRV-2}. More precisely,
\be
f_\chi(\widetilde{t}) = \, \mu_\chi\, F_\chi(s \widetilde{t}),
\qquad\qquad 
f_\xi(\widetilde{t}) = \, \mu_\xi\, F_\xi(s \widetilde{t}),
\label{eq4.4}
\ee
for appropriate constants $\mu_\chi$, $\mu_\xi$, and $s$. These relations
are shown rigorously to all orders in the expansion around 
the mean-field limit in Ref. \cite{PRV-2} and provide the 
link between medium-range crossover functions and field-theoretical 
expressions.
The constants can be easily computed by comparing the behavior 
for $\widetilde{t}\to \infty$. In three dimensions, 
the functions $ F_\chi(\widetilde{t}_{\rm SR})$
and $F_\xi(\widetilde{t}_{\rm SR})$ behave as (see App. \ref{secA.1.1})
\begin{eqnarray}
F_\chi(\widetilde{t}) &=&
   {1\over \widetilde{t}}\left(1 + {1\over 12\pi} \widetilde{t}^{-1/2}  
      +\, O(\widetilde{t}^{-1}\log \widetilde{t})\right), 
\nonumber \\
F_\xi(\widetilde{t}) &=&  
   {1\over \widetilde{t}}\left(1 + {1\over 12\pi} \widetilde{t}^{-1/2}  
      +\, O(\widetilde{t}^{-1}\log \widetilde{t})\right).
\label{fchiximf}
\eea
By comparing these expansions with Eqs. \reff{fchimf} and \reff{fximf},
we obtain
\be
s = \mu_\chi = \mu_\xi = {1\over 9}.
\label{rescaling-3d}
\ee
We can now use the explicit results of App. \ref{AppB} to obtain predictions
for the constants $A_\chi^\infty$, $A_\xi^\infty$, $B_\chi^\infty$, and
$B_\xi^\infty$ defined in Eqs. \reff{fchi-Wilson-Fisher} and
\reff{fxi-Wilson-Fisher}. We obtain in the fixed-dimension expansion 
(see App. \ref{secA.1.2})
\bea
A_\chi^\infty &=& \mu_\chi\, \chi_0 s^{-\gamma} = 0.5959 \pm 0.0041,
\\
A_\xi^\infty  &=& \mu_\xi\, \xi^2_0 s^{-2 \nu} = 0.5238 \pm 0.0024,
\\
B_\chi^\infty &=& \chi_1 s^\Delta = 2.18 \pm 0.18,
\\
B_\xi^\infty &=& \xi_1 s^\Delta = 2.92 \pm 0.27.
\eea
For the (universal) ratio $B_\chi^\infty/B_\xi^\infty$ we obtain the 
more precise result
\be
{B_\chi^\infty\over B_\xi^\infty} =\, 0.745 \pm 0.034.
\ee
Consistent, although less precise, results can be obtained in the 
framework of dimensional regularization without $\epsilon$-expansion,
see App. \ref{secA.2.2}.

In a completely analogous way we can derive from field theory the 
crossover functions $g_c(\widetilde{n})$ and $g_E(\widetilde{n})$. 
Indeed, we introduce functions $G_c(\widetilde{n}_{\rm SR})$ and 
$G_E(\widetilde{n}_{\rm SR})$ in the following way:
\begin{eqnarray}
F_\chi(t) &=& \int^\infty_0 du\,G_c(u) e^{-ut},
\label{Fchi-Gc}  \\
F_\xi(t) F_\chi(t) &=& {1\over 2d} \int^\infty_0 du\, G_c(u) G_E(u) e^{-ut}.
\label{Fxi-GE}
\end{eqnarray}
The functions $G_c(\widetilde{n}_{\rm SR})$ and 
$G_E(\widetilde{n}_{\rm SR})$ can be computed perturbatively
by using the corresponding perturbative expressions for 
$F_\chi(\widetilde{t}_{\rm SR})$ and $F_\xi(\widetilde{t}_{\rm SR})$.
The relevant formulae are reported in App. \ref{secA.3}.

In the fixed-dimension expansion, using the seven-loop results of 
Ref. \cite{Murray-Nickel_91}, we obtain (the six-loop result for 
$G_E(\widetilde{n}_{\rm SR})$ already appears in 
Ref. \cite{Muthukumar-Nickel_87})
\bea
G_c(\widetilde{n}_{\rm SR}) &=& e^{4 \pi z^2 \log (Kz)}\left[
    1 + 4 z + 2 \pi \gamma_E z^2 - 60.7295 z^3 -
    96.6721 z^4 \right.
\nonumber \\
   && \qquad \left. - 144.431 z^5 + 2491.95 z^6 - 5070.31 z^7 + O(z^8)\right],
\label{Gcpert}
\\
G_E(\widetilde{n}_{\rm SR}) &=& 6 \widetilde{n}_{\rm SR} \left[
1 + {4\over3} z + \left({28\pi\over 27} - {16\over 3}\right) z^2 +
         6.29688 z^3 - 25.0573 z^4 + \right. 
\nonumber \\
         && \qquad \left. 116.135  z^5 - 594.717   z^6 + 3273.16 z^7 + 
        O(z^8) \right],
\label{GEpert}
\end{eqnarray}
where $K$ is a nonperturbative constant and
\be
z = {1\over 24 \pi} \left({\widetilde{n}_{\rm SR}\over \pi}\right)^{1/2}.
\ee
Explicitly
\be
\log K = 144 \pi^2 D_3 + {1\over2} \log (16 \pi) - {34\over 9},
\ee
where $D_3$ is a nonperturbative constant reported in App. \ref{secA.1.1}.
Numerically, using the estimate of $D_3$ reported in App. \ref{secA.1.1},
we have $K = 5.44(5)$.

These perturbative expressions can be resummed by using the fact 
that the series are Borel summable \cite{EMS-75,FO-76,MS-77,MR-85}.  
The technical details are 
reported in App. \ref{secA.3}. The resummation is very precise for 
$z\ltapprox 1$, with errors smaller than 0.2\%. For larger values of 
$z$ the resummation errors increase and the numerical integration becomes
unstable: In practice we have not been able to compute numerically
the crossover functions using Eqs. \reff{Gcfinale}, \reff{hEfinale} 
for $z\gtapprox 5$. However, in this region the crossover functions are 
already well approximated by the asymptotic expansions 
\reff{hcztoinfty}, \reff{hEztoinfty}. The resummed expressions are 
well fitted by the following simple formulae:
\bea
G_c(\widetilde{n}_{\rm SR}) &=& 
         (1 + 50.79365 z + 508.5428 z^2 + 
              5929.475 z^3 + 10937.03 z^4)^{0.07875},
\label{Gcresum}
\\
G_E(\widetilde{n}_{\rm SR}) &=& 6 \widetilde{n}_{\rm SR} 
    (1 + 7.6118 z + 12.05135 z^2)^{0.175166}.
\label{GEresum}
\eea
The expression for $G_E(\widetilde{n}_{\rm SR})$ was proposed in Ref.
\cite{Belohorec-Nickel_97} and it was obtained from a detailed 
Monte Carlo study of the Domb-Joyce model.\footnote{Other representations
\cite{Stockmayer,YT-67,DB-76,desCloizeaux-81,DF-84-85,dCCJ-85} 
for $G_E(\widetilde{n}_{\rm SR})$ are reviewed in Ref. 
\cite{Muthukumar-Nickel_87}. Note that in Ref. \cite{Muthukumar-Nickel_87} 
$\alpha^2$ stands for $G_E(\widetilde{n}_{\rm SR})/(6 \widetilde{n}_{\rm
SR})$.} We find that the perturbative 
results are very well described by these expressions, with discrepancies 
of less than 0.3\% for $z<2$. For larger values of $z$ differences are 
slightly larger, of the order of 1\%, which is, in any case, of the same
order of the error of our resummed results.
Note that the expressions \reff{Gcresum} and \reff{GEresum} exactly reproduce 
the small-$z$ behaviors \reff{Gcpert} and \reff{GEpert} 
up to terms of order $O(z^2)$.

The relation between the field-theory functions and $g_c(\widetilde{n})$,
and $g_E(\widetilde{n})$ is straightforward. From 
Eqs. \reff{eq4.4} and \reff{rescaling-3d} we have
\be
g_c(\widetilde{n}) =\, \lambda_c\, G_c(\rho \widetilde{n}), \qquad \qquad
g_E(\widetilde{n}) =\, \lambda_E\, G_c(\rho \widetilde{n}),
\label{relG}
\ee
with
\be
\lambda_E = {1\over9}, \qquad \lambda_c = 1, \qquad \rho = 9.
\label{relconst}
\ee
Using the results of App. \ref{secA.3} we can easily derive estimates for 
the constants $C_\chi^\infty$, $C_E^\infty$, 
$D_\chi^\infty$, and $D_E^\infty$ defined in Eqs. \reff{gc-WF}, \reff{gE-WF}.
In the fixed-dimension expansion we have 
\bea 
C_\chi^\infty &=& 0.640 \pm 0.005,\\
C_E^\infty    &=& 2.457 \pm 0.011,\\
D_\chi^\infty &=& 1.45 \pm 0.10,\\
D_E^\infty    &=& 5.03 \pm 0.48. 
\eea
If we consider the universal ratio $D_\chi^\infty/D_E^\infty $ we 
obtain the more precise result
\be
{D_\chi^\infty\over D_E^\infty  } = 0.288 \pm 0.016.
\ee
We mention that from the very precise Monte Carlo results 
of Ref. \cite{Belohorec-Nickel_97} we would obtain 
$C_E^\infty\approx 2.450$ and $D_E^\infty \approx 5.57$, 
in reasonable agreement with our results.

\section{Algorithms} \label{sec5}

The SAW with nearest-neighbor jumps can be very efficiently simulated 
by means of nonlocal algorithms \cite{Madras-Sokal_88,CPS-92}.
None of them can be generalized to the 
case at hand, and thus we have resorted to the dimerization algorithm (DA)
\cite{Alexandrowicz_69,Alexandrowicz-Accad_71}. Although
the CPU time needed to generate a walk increases more than any power of its
length \cite{Madras-Sokal_88}, the prefactors are so small that 
we can reach quite large lengths. It should be noticed that 
other algorithms could have probably performed better.  For instance, we could 
have used the pruned-enriched Rosenbluth method of Ref.
\cite{Grassberger-95-97}.

Before defining the DA, let us introduce the simple-sampling
algorithm (SSA).
The SSA is the simplest algorithm for the  generation of  SAWs. 
It builds a walk recursively. Once an $n$-step 
$R$-SAW $\{\omega_0\ldots,\omega_n\}$ is 
generated, an $(n+1)$-step $R$-SAW is obtained by choosing at random a 
new point $\omega_{n+1}$ in $D_{R}(\omega_n)\backslash\{\omega_n\}$. If 
the new walk is self-avoiding it is kept, otherwise the 
$n$-step $R$-SAW is discarded and the procedure starts again from scratch
generating a new $n$-step $R$-SAW. Since adding one step and 
checking for self-avoidance requires\footnote{The self-avoidance 
check can be performed in a CPU-time of order one by using 
a hashing technique, see, e.g., Refs. \cite{Madras-Sokal_88,Binder_SV}.}
$O(1)$ operations, the 
CPU time needed to generate an $n$-step walk is 
\be
T_{\rm CPU}(n) \approx {\beta_{c,\rm mf}^{-n}\over c_n} 
        \sum_{m=1}^n {c_m \beta_{c,\rm mf}^m},
\ee
where $\beta_{c,\rm mf}\equiv 1/(V_R-1)$.
In the limit $n\to\infty$ with $R$ fixed and large, using Eqs. 
\reff{cnRasym} and \reff{betacR-3d}, 
we obtain
\be
T_{\rm CPU}(n)\sim R^{d \gamma} n^{-\gamma + 1} e^{\alpha n R^{-d}},
\ee
where $\alpha$ is defined by 
$\beta_{c,\rm mf}/\beta_{c,R}\approx 1 - \alpha R^{-d}$. For our model
$\alpha \approx 0.035$.
The computer time increases exponentially with $n$ although the factor in the 
exponential goes to zero as $R^{-d}$. 

The SSA is quite efficient in generating short walks. However,
far from the Gaussian region it becomes too slow, because of the 
exponentially increasing time needed to generate a walk. A better algorithm 
is the DA \cite{Alexandrowicz_69,Alexandrowicz-Accad_71}. 
Numerically, we find DA to perform better
than SSA for $n\gtapprox V_R$. The DA is again a recursive 
algorithm. To generate an $n$-step walk one generates two $n/2$-step
walks and concatenates them. If the resulting walk is self-avoiding, it 
is kept, otherwise the two $n/2$-step walks are discarded and the 
procedure is repeated again. The algorithm is recursive: in order to 
generate the walks of length $n/2$, the DA is used again until
$n/2 < n_c$. If $n/2<n_c$, we generated the walks using the SSA. 
In our implementation we chose $n_c\approx V_R$. 
The behavior of the DA in the limit $n\to\infty$ at $R$ fixed
was studied in Ref. \cite{Madras-Sokal_88}. By using the results of 
Sec. \ref{sec3}, one finds 
\be
T_{\rm CPU}(n) \sim R^{q_1} n^{q_2} 
    \exp \left[ {(\gamma - 1)\over 2\log2} 
         \log^2 \left(n R^{2 d (1 - \gamma)/(4-d)}\right)\right],
\ee
where $q_1$ and $q_2$ are exponents that depend on the specific model
and on the implementation of the algorithm.

Let us now discuss how to estimate $E^2_{n,R}$ and $c_{n,R}$ from the 
simulation. 
Estimating $E^2_{n,R}$ is completely straightforward. 
To estimate $c_{n,R}$ we have used the acceptance fraction for the 
elementary moves of the two algorithms. Indeed, given 
an $n$-step $R$-SAW, the probability of obtaining 
an $(n+1)$-step walk using the SSA is simply $c_{n+1} \beta_{c,\rm mf}/c_n$. 
Thus, if we know in a given SSA simulation the number $N_n$ of 
generated walks of length $n$,we can 
estimate $c_n$ using the recursion relation
\be
c_n = c_{n-1} \beta_{c,\rm mf}^{-1}\, {N_n\over N_{n-1}},
\label{recforcnSSA}
\ee
with the initial condition $c_1 = \beta_{c,\rm mf}^{-1}$.

Analogously, given two $R$-SAWs of length $n$, the probability 
that their concatenation is an $R$-SAW is simply $c_{2n}/c_{n}^2$. 
Therefore, if we know in a given DA simulation the number 
$N_n$ of generated walks of length $n$, we can compute $c_n$
using 
\be
c_{n} = c^2_{n/2} {N_n\over N_{n/2}}
\ee
for $n\ge n_c$ and then Eq. \reff{recforcnSSA}.

Note that in a dimerization simulation in which we generate walks 
of maximal length $n_{\rm max}$, we obtain at the same time estimates 
of the observables also for a set of smaller values of $n$, i.e. 
for $n = n_{\rm max}/2$, $n_{\rm max}/4$, $\ldots$ These results 
are of course correlated, especially in the mean-field region where the 
rejection rate at each step is small. However, for our global observables,
the correlation should be small. Analogously, when we use the 
SSA, we can compute the observables for all values of $n$, although the 
results are in this case strongly correlated.

\section{Numerical results} \label{sec6}

We have performed an extensive simulation using three-dimensional 
walks with $n\le 66560$ and $2\le \rho \le 12$. Notice 
that the values of $\rho$ are particularly large: for 
$\rho = 12$, in the spin language, each spin interacts 
with 2624 neighbors. The advantage of working with 
SAWs is the absence of finite-size effects---we work in the 
infinite-volume limit---and the possibility of reaching 
large values of the correlation length.\footnote{In order to compare 
with spin models,
it is interesting to relate our values of $E^2_{n,R}$ with $\xi_R(\beta)$, 
where, for each $n$, we consider the value of $\beta$ such that 
$\< n\>_\beta = n$. In the critical limit we find the relation
\[
\xi_R(\beta)^2 = {\Gamma(\gamma + 2\nu) \over 6 \Gamma(\gamma)
                 \gamma^{2\nu}}\ E^2_{n,R}. 
\]
Therefore, we reach $\xi \approx 35,80,300,500,350,500,500,700,700,500$, 
respectively for $R=2,3,4,5,6,7,8,9,10,12$. } 
The raw data for the largest values\footnote{For $\rho=2$, 7, 12, 
we have computed the observables for additional values of $n$, larger than
the minimum value of $n$ reported in the Table. More precisely,
by using the SSA, we have computed the observables for all 
$n\le 30$, 400, 3000 respectively. }
of $n$ and several values of 
$R$ are reported in Tables \ref{tablerawdata} and  \ref{tablerawdata2}.

In Sec. \ref{sec6.1} we will determine $\beta_{c,R}$ from our numerical 
data and we will explicitly check the theoretical predictions for the 
large-$R$ behavior of $\beta_{c,R}$ of Ref. \cite{PRV-2} presented 
in Sec. \ref{sec3.1}. In Sec. \ref{sec6.2} we will compute the critical 
crossover functions and we will compare them with the field-theoretical 
results of Sec. \ref{sec4}.

\subsection{Determination of $\beta_{c,R}$} \label{sec6.1}

In order to compute $\beta_{c,R}$ we define
\be
\beta_{{\rm eff},R}(n) \equiv \left[ {c_{n,R}\over g_{c,\rm th}(\widetilde{n})}
        \right]^{-1/n},
\label{betaeff-def}
\ee
where $g_{c,\rm th}(\widetilde{n})$ is the theoretical crossover 
function: in our numerical determination of 
$\beta_{{\rm eff},R}(n)$ we will use\footnote{In principle, we could have used 
Eq. \reff{Gcresum} for $G_c(z)$. However, here it is important to use 
an expression for $g_{c,\rm th}(\widetilde{n})$ which is 
precise in the Wilson-Fisher region. For this reason, we have chosen
to use Eq. \reff{GcresumApp}.} 
Eqs. \reff{relG} and \reff{GcresumApp}. 
By using Eqs. \reff{cnRasym} and \reff{gc-WF}, we obtain for $n\to\infty$
\be
\beta_{{\rm eff},R}(n) = 
  \beta_{c,R} \left[1 - {1\over n} \log\left({C_\chi(R) R^{6(\gamma-1)}\over
         C_\chi^\infty}\right) + {1\over n^{1+\Delta}} 
       (D_\chi(R) - D_\chi^\infty R^{6\Delta}) + \cdots\right].
\ee
Using the asymptotic expansions \reff{eq3.4} and the relations 
\reff{relazioni-ampiezze}, we obtain for $R\to \infty$
\be
\beta_{{\rm eff},R}(n) = 
      \beta_{c,R} \left[1 + {\alpha_1\over n R^3} +  
    {\alpha_2 R^{6\Delta-3} \over n^{1+\Delta}} + \cdots\right],
\ee
where $\alpha_1$ and $\alpha_2$ are $R$-independent constants.
This expression shows the advantage of the definition \reff{betaeff-def}
over the common one in which one simply considers $(c_n)^{-1/n}$. 
Indeed, with our choice, the $1/n$ correction vanishes for $R\to \infty$
while the $1/n^{1+\Delta}$ remains approximately constant ($\Delta\approx
1/2$); with the other one, we would have corrections of order 
$\log (n R^{-6})/n$ and $R^{6\Delta}/n^{1+\Delta}$.
This improved behavior is particularly important, since for large 
$R$ we are quite far from the Wilson-Fisher point, and thus a reduction 
of the scaling corrections is essential in order to obtain precise 
estimates of $\beta_{c,R}$.
In order to determine $\beta_{c,R}$ 
we have performed fits of the form\footnote{In the fits we also used 
estimates obtained by means of the SSA. In order to reduce the 
correlations and to avoid the fit to be dominated by data with small 
values of $n$, we have used only some of them. We considered 
the values corresponding to $n = n_0 \lfloor e^{(k+k_0)/5}\rfloor$, $k
\in N$, where $\lfloor x \rfloor$ is the largest integer smaller than $x$ and
$n_0$ and $k_0$ are integers.}
\be
\beta_{{\rm eff},R}(n) V_R = \beta_{c,R} V_R + 
         {a\over n} + {b R^{6\Delta}\over n^{1+\Delta}},
\label{fitbetac}
\ee
assuming $\Delta =1/2$. We have repeated the fit several times, considering
each time only the data satisfying $n\ge n_{\rm min}$. The final results,
reported in Table \ref{tablebetac}, correspond to the smallest
$n_{\rm min}$ for which $\chi^2/{\rm d.o.f.}\approx 1$ (d.o.f. is the number 
of degrees of freedom). Notice that, by rescaling $b$ by $R^3$, the
coefficients $a$ and $b$ should become $R$-independent as $R$ increases. 
This is evident for $a$ and indeed we can roughly estimate 
$a\approx -0.015(5)$ for $R\to\infty$. This allows us to compute 
the leading correction to $C_\chi(R)$.  We have 
\be 
C_\chi(R) \approx C_\chi^\infty R^{6(1-\gamma)} (1 + {\kappa}_\chi R^{-3} + 
  \cdots), 
\ee
where ${\kappa}_\chi = - a $. The results for 
$b$ are less stable, but still reasonably compatible with a constant 
for large $R$.
In order to understand the systematic errors due to 
the truncation \reff{fitbetac} we have repeated the fit with an additional
correction:
\be
\beta_{{\rm eff},R}(n) V_R = \beta_{c,R} V_R + 
         {a\over n} + {b R^{6\Delta}\over n^{1+\Delta}} + 
     {c R^6\over n^{2}}.
\label{fitbetac22}
\ee
The results for $\beta_{c,R}$ do not differ significantly from those of
the fit with $c=0$, except for $\rho = 12$, where the difference is 
approximately three combined error bars.
Therefore, our final 
estimates should be quite reliable. As an additional check we have compared 
our results with the theoretical prediction \reff{betacR-3d}. 
If we indicate with $\beta^{\rm (exp)}_{c,R}$ the expansion \reff{betacR-3d}
neglecting terms of order $R^{-8}$, we define 
\be
\tau_{2,\rm eff} \equiv R^8 \left({\beta_{c,R}\over \beta^{\rm (exp)}_{c,R}}
       -1 \right).
\ee
If we have correctly determined $\beta_{c,R}$, $\tau_{2,\rm eff}$ 
should converge to the constant $\tau_2$ as $R\to \infty$, with corrections 
of order $\log R^2/R$. The plot of $\tau_{2,\rm eff}$ is reported in
Fig. \ref{tau2}. A fit of the form $\tau_{2,\rm eff} = \tau_2 + b/R$ gives
\be
\tau_2 = - 0.00814(21),
\ee
$b = 0.00277(19)$, including all data with $\rho \ge 3$.

\subsection{Determination of the critical crossover functions} \label{sec6.2}

We wish now to determine the critical crossover functions in three dimensions. 
We begin by studying the function
$\widetilde{c}_{n,R}$, cf. Eq. \reff{defctilden}. The function is reported 
in Fig. \ref{CNscal1} (upper graph) together with the theoretical 
prediction obtained by using  Eqs. \reff{relG} and
\reff{Gcresum}. Note that there is no free parameter in the 
theoretical curve. We observe a very good agreement, especially in the 
Wilson-Fisher region. Systematic deviations are observed for smaller
values of $\widetilde{n} $. In order to understand the role of the 
deviations we report in Fig. \ref{CNscal1} (lower graph) the same data, 
but now we exclude 
all points with $n < V_R/2$. The agreement is now perfect for all 
$\rho \ge 3$. We thus clearly see that the crossover behavior requires 
$n\gg R^3$. In particular, mean-field behavior is always observed 
for $n\ll R^{6}$ if $R$ is sufficiently large, but it is 
described by the critical crossover curves only if 
$R^3\ll n  \ll R^{6} $.

In order to see the corrections to scaling, in Fig. \ref{CNscal2} we report 
the ratio $\widetilde{c}_{n,R}/g_{c,\rm th}(\widetilde{n})$ which should 
converge to 1 as $R\to\infty$. Corrections to scaling 
are clearly evident, points with different values of $R$ lying 
on different curves that indeed converge to 1 as $R\to\infty$. 
These corrections are predicted to scale as $R^{-d}$. To check this behavior 
we considered 
\be
\Delta_{c;n,R} \equiv R^3 
  \left( {\widetilde{c}_{n,R}\over g_{c,\rm th}(\widetilde{n})} - 1\right),
\ee
which should converge to $k_c(\widetilde{n})/g_c(\widetilde{n})$ 
in the crossover limit. Using the expected asymptotic behavior of 
$k_c(\widetilde{n})$ and $g_c(\widetilde{n})$, 
$\Delta_{c;n,R} $ converges to a constant both for 
$\widetilde{n}\to 0$ and $\widetilde{n}\to \infty$. For 
$\widetilde{n}\to 0$, $k_c(0)/g_c(0) =  - E_3 \approx - 0.059$.

The numerical results are reported in Fig. \ref{CNscal3}, where the error on 
$\Delta_{c;n,R}$ has been computed by considering the error on 
$\beta_{c,R}$, $c_{n,R}$, and on the theoretical curve.\footnote{
We assume $g_{c,th}(\widetilde{n})$ to have an error of 0.3\% 
for $\widetilde{n}\ge 1$, and of $0.3\widetilde{n}$ \% for 
$\widetilde{n} < 1$. The smaller error for $\widetilde{n} < 1$ 
is due to the fact that our representation \reff{Gcresum} has the 
exact leading behavior for $z\to 0$, cf. Sec. \ref{sec4}.} 
A reasonably good scaling is observed, confirming the results of 
Ref. \cite{PRV-2}. Also the prediction $k_c(0)/g_c(0)\approx - 0.059$ 
is fully compatible with the data.

In order to perform a more precise check, we have also considered 
the quantity 
\be 
Q_{n,R} \equiv {c^2_{n,R}\over c_{2n,R}},
\ee
that 
converges to $g_c(\widetilde{n})^2/g_c(2\widetilde{n})$ in the critical
crossover limit. For $Q_{n,R}$ we do not need the value of $\beta_{c,R}$ 
and thus a source of error is avoided. In Fig. \ref{Qscal} we 
show $Q_{n,R}$ together with the theoretical prediction and 
\be
\Delta_{Q;n,R} \equiv R^3 \left( {Q_{n,R}\,  g_{c,th}(2 \widetilde{n}) \over 
        g_{c,th}( \widetilde{n})^2 } - 1\right).
\ee
The errors on $\Delta_{Q;n,R}$ have been computed as we did for 
$\Delta_{c;n,R}$.
The agreement between the numerical data and the theoretical 
prediction is very good. Also, $\Delta_{Q;n,R}$ shows a nice scaling 
behavior confirming that the corrections scale as $R^{-d}$. 

Let us finally discuss the effective exponent 
$\gamma_{\rm eff}(n,R)$. A standard
definition would be
\be
\gamma_{\rm eff} (n,R) \equiv 1 + n {d\log \widetilde{c}_{n,R}\over dn}.
\ee
However, this definition is not easy to use in numerical simulations since it
involves the derivative with respect to $n$. Here, we will use the  
definition
\bea
\gamma_{\rm eff} (n,R) \equiv\;  1 + {1\over \log 2} \,
    \log\left({\widetilde{c}_{2n,R}\over \widetilde{c}_{n,R}}\right),
\eea
which interpolates between the SAW value
$\gamma = 1.1575$ and the mean-field value $\gamma = 1$. 
The results are reported in Fig. \ref{gammaeff} 
together with the theoretical prediction.  The agreement is very good. 
Note that in the Wilson-Fisher region the numerical data are 
well approximated by the field-theoretical prediction only for 
$R\gtapprox 4$. For smaller values of $R$, the corrections are important,
as it was already noticed in the Ising model simulations 
\cite{Luijten-Binder_98,Luijten}.

The analysis we have performed for $c_n$ can be repeated for $E^2_{n,R}$.
In this case however the errors are smaller since the critical crossover
functions do not depend on $\beta_{c,R}$.
In Fig. \ref{Escal} (upper graph) we report our results for 
$\widetilde{E}^2_{n,R}$ together
with the prediction $g_{E,\rm th}(\widetilde{n})$ 
obtained by using the field-theory result \reff{GEresum} and the 
relations \reff{relG}, \reff{relconst}. Note that there is no 
free parameter in the theoretical curve. 
The agreement is very good although one can see clearly the presence 
of corrections to scaling.

We wish now to compute the correction curve $k_E(\widetilde{n})$. For this
purpose we consider
\be
\Delta_{E;n,R}\equiv R^3
   \left({\widetilde{E}^2_{n,R} \over g_{E,\rm th}(\widetilde{n})} - 1\right),
\ee
that converges to $k_E(\widetilde{n})/g_E(\widetilde{n})$ as $R\to\infty$. 
The plot of $\Delta_{E;n,R}$ is reported in Fig. \ref{Escal} (lower 
graph), where we have taken only into account the error on 
$\widetilde{E}^2_{n,R}$. A good scaling
behavior is observed confirming the theoretical prediction for the 
corrections. Moreover, this nice scaling behavior is also an 
indication that the approximation \reff{GEresum} can be considered 
at our level of precision practically exact.
Note also that the prediction $k_E(0)/g_E(0) = - 0.059$, cf. 
Eq. \reff{corrections-k}, is in good agreement with our data.

Let us finally discuss the effective exponent $\nu_{\rm eff}$. As we did for 
$\gamma_{\rm eff}$, instead of 
\be
\nu_{\rm eff} (n,R) \equiv n {d\log E^2_{n,R}\over dn},
\ee
we will consider
\be
\nu_{\rm eff} (n,R) \equiv\;  {1\over 2\log 2} \,
    \log\left({{E}^2_{2n,R}\over E^2_{n,R}}\right),
\ee
that is easier to compute numerically.
A graph of this quantity is reported in Fig. \ref{nueff}. It shows the
 expected 
behavior: for $\widetilde{n} \to 0$ it converges to 1/2, while
for $\widetilde{n} \to \infty$ it converges to $\nu_{SAW} \approx 0.588$.
The agreement with the perturbative prediction is quite good
in the random-walk region. On the other hand, as $\widetilde{n}$ increases
the corrections increase, in agreeement with similar results obtained
for the Ising model \cite{Luijten-Binder_98,Luijten}.

\section{Conclusions} \label{sec7}

In this paper we have studied the critical crossover limit for a model 
of walks and we have verified numerically the following statements,
predicted by field-theoretical and mean-field methods \cite{PRV-2}:
\begin{itemize}
\item[1)] The critical crossover functions in medium-range models 
{\em coincide} with the field-theoretical crossover curves. The nonuniversal 
constants can be determined by computing the corrections to the 
mean-field behavior.
\item[2)] The asymptotic behavior of $\beta_{c,R}$ for $R\to \infty$ 
can be computed exactly in lattice models up to corrections of relative order 
$R^{-8}$ by determining the first corrections to the mean-field limit 
and exploiting the field-theoretical model.
\item[3)] The corrections to the critical crossover functions decrease 
as $R^{-d}$ once $R$ is defined as in Eq. \reff{defR2}.
\end{itemize}
Our numerical results can also be used to determine phenomenological expressions
that describe the data for all values of $R$. Here, different procedures
can be used. One can consider the phenomenological model of Ref. 
\cite{ALASB-99} (with the modifications discussed in Ref. \cite{PRV-2} 
to make it compatible with the theoretical predictions) or use the procedure
proposed in Ref. \cite{PRV-2}. The idea is to write 
\bea
\widetilde{c}_{n,R} = g_c(\widetilde{n}) + R^{-3} k_c(\widetilde{n}), 
\nonumber \\
\widetilde{E}^2_{n,R} = g_E(\widetilde{n}) + R^{-3} k_E(\widetilde{n}), 
\label{phenomenological1}
\eea
and use a simple parametrization for the correction terms. Here, we approximate
\be
k (\widetilde{n}) = g(\widetilde{n}) 
   {-0.059 + a \widetilde{n}^{1/2} + b \widetilde{n} \over 
    1 + c \widetilde{n}^{1/2} + d \widetilde{n} }
\label{phenomenological2}
\ee
both for $k_c$ and $k_E$, where we have used the asymptotic behavior 
\reff{corrections-k}. The parameters $a$, $b$, $c$, and $d$ are 
determined by fitting the numerical data. The best results are obtained for 
$a=-61$, $b=-1.06$, $c=1830$, $d=87$ (function $k_c$)
and $a = -23$, $b = 0.8505$, $c = 972$, $d=32$ (function $k_E$).
These fitting functions provide phenomenological expressions that correctly
describe our data for all values of $\rho$. 
The corresponding effective exponents are reported in Fig. \ref{exp-phen}
and show the typical behavior that has been found in simulations of the 
Ising model. We have also included in the figure the curves corresponding
to $\rho =3/2$, to show that for $\rho$ small the phenomenological 
expressions show a nonmonotonic behavior that is not present in the 
critical crossover curve.

Although our main motivation was the understanding of spin models, the results 
of this paper are also relevant in the context of polymers. 
Indeed, as is well known \cite{deGennes_book,desCloizeaux-Jannink_book},
noninteracting SAWs describe the universal behavior of homopolymers 
in dilute solutions above the $\Theta$-temperature. In the polymer 
context, however, it is more interesting to consider a 
different model with medium-range interactions. Supposing, for 
simplicity, to be in the continuum (off-lattice), we can define 
a SAW in the following way. A SAW with medium range interactions is a 
collection of point $\{\omega_0,\ldots,\omega_n\}$, $\omega_i \in R^d$, 
such that $|\omega_i - \omega_{i+1}| = \rho$ and $|\omega_i - \omega_j|>a$
for all $i\not= j$. In this case the relevant scale is $\rho/a$ and the 
crossover limit is obtained for $\rho/a\to\infty$, $n\to\infty$, with 
$\widetilde{n} \equiv n (\rho/a)^{-2d/(4-d)}$ fixed. For this 
model, the critical crossover functions can also be computed using 
Eq. \reff{relG}, although with different nonuniversal $\rho$-independent 
constants $\lambda_E$, $\lambda_c$, and $\rho$. Thus, the results presented 
here are relevant for the description of polymeric systems in which the 
macromolecular persistence  is much larger than the molecular scale. In practice,
we expect the description to be reasonably accurate when $\rho/a\gtapprox 3$.


\appendix

\section{Critical crossover functions from field theory} \label{AppB}

In this Appendix we will compute the critical crossover functions for the 
polymer case using field-theory methods. In Section 
\ref{secA.1} we will use the approach of 
Refs. \cite{Bagnuls-Bervillier_84,Bagnuls-Bervillier_85},
while in Section \ref{secA.2} we will present the results obtained 
using the method of Refs. \cite{Dohm,Schloms-Dohm_89,K-S-D}. 
The first approach provides the most precise estimates and it will
be applied in Sect. \ref{secA.3} to obtain numerical results 
for polymers generalizing Ref. \cite{Muthukumar-Nickel_87}.

\subsection{Crossover functions in the fixed-dimension expansion} \label{secA.1}

\subsubsection{General results} \label{secA.1.1}

In this Section we report the 
critical crossover functions using the approach of 
Refs. \cite{Bagnuls-Bervillier_84,Bagnuls-Bervillier_85}. We start from 
the expressions for
$F_\chi(\widetilde{t})$ and $F_\xi(\widetilde{t})$:
\begin{eqnarray}
F_\chi(\widetilde{t}) &=& \chi^* 
   \, \exp\left[ - \int_{y_0}^g dx\, {\gamma(x)\over \nu(x) W(x)}\right], 
\label{Fchi-FT-fixedd}\\
F_\xi(\widetilde{t}) &=& \left(\xi^*\right)^2
   \, \exp\left[ - 2 \int_{y_0}^g dx\, {1\over W(x)}\right], 
\label{Fxi-FT-fixedd}
\end{eqnarray}
where $\widetilde{t}$ is related to the zero-momentum four-point
renormalized coupling $g$ by
\be
\widetilde{t} \, =\, 
  - t_0 \, \int^{g^*}_g dx\, 
 {\gamma(x)\over \nu(x) W(x)} 
  \exp\left[ \int_{y_0}^x dz\, {1\over \nu(z) W(z)} \right],
\label{ttilde-FT-fixedd}
\ee
$\gamma(x)$, $\nu(x)$, and $W(x)$ are the standard renormalization-group (RG)
functions, $g^*$ is the critical value of $g$ defined by
$W(g^*) = 0$, and $\chi^*$, $\xi^*$, $t_0$, and $y_0$ are normalization
constants. 

The expressions \reff{Fchi-FT-fixedd}, \reff{Fxi-FT-fixedd}, and 
\reff{ttilde-FT-fixedd} are valid for any dimension $d<4$. The first
two equations are always well defined, while Eq. \reff{ttilde-FT-fixedd}
has been obtained with the additional hypothesis that the integral
over $x$ is well defined when the integration is extended up to 
$g^*$. This hypothesis is verified when the system becomes critical 
at a finite value of $\beta$ and shows a standard critical behavior. 
In our case, $N=0$, this is true for all $1\le d < 4$.

We normalize the coupling $g$ as in Refs.
\cite{Baker-etal_77_78,LeGuillou-ZinnJustin_80} so that in the perturbative
limit $g\to 0$, $t \to \infty$, we have
\be
g \approx \, 
  {4\over 3 (4\pi)^{d/2}} \Gamma\left(2 - {d\over2}\right)\,
    \widetilde{t}^{(d-4)/2} \, \equiv \lambda_d \widetilde{t}^{(d-4)/2}
\ee
This implies that for $y_0\to 0$ we have 
$t_0 \approx (y_0/\lambda_d)^{2/(d-4)}$ and 
$(\xi^*)^2 t_0 \approx \chi^* t_0 \approx 1$. 
With this normalization, the previous equations can be written as
\begin{eqnarray}
F_\chi(\widetilde{t}) &=& (6 \pi g)^2\, 
   \, \exp\left[ - \int_{0}^g dx\, 
    \left({\gamma(x)\over \nu(x) W(x)} + {2\over x}\right) \right], 
\label{Fchi-FT-fixedd2}\\
F_\xi(\widetilde{t}) &=& (6 \pi g)^2\,
   \, \exp\left[ - 2 \int_{0}^g dx\, 
     \left({1\over W(x)} + {1\over x}\right) \right], 
\label{Fxi-FT-fixedd2}
\end{eqnarray}
and
\begin{eqnarray}
&&\widetilde{t} = {1\over (6\pi g)^2} 
    \left(1 - {3 g\over 2} + {g^2\over 4}\log g\right) + D_3 
\\
&& +  {1\over (6\pi)^2}\, \int_0^{g} {dx\over x^2}
    \left\{ {\gamma(x)\over \nu(x) W(x)}
       \exp\left[\int_0^x dz\, \left({1\over \nu(z) W(z)} + {2\over z}\right)
           \right]\, 
      + {2\over x} - {3\over 2} - {x\over4} \right\},
\nonumber
\eea
where $D_3$ is a nonperturbative constant given by 
\bea
&&D_3 =
- {1\over (6\pi)^2}\,
    \left[{1\over (g^{*})^2} - {3\over 2 g^*} +
     {1\over4} \log g^*\right]
\\
&& - {1\over (6\pi)^2}\, \int_0^{g^*} {dx\over x^2}
    \left\{ {\gamma(x)\over \nu(x) W(x)}
       \exp\left[\int_0^x dz\, \left({1\over \nu(z) W(z)} + {2\over z}\right)
           \right]\, 
      + {2\over x} - {3\over 2} - {x\over4} \right\}.
\nonumber
\eea
Numerically, $D_3 = 0.002473(6)$. Expressions for a general $N$-vector 
model and in two dimensions can be found in Ref. \cite{PRV-2}.

For $\widetilde{t}\to \infty$, in three dimensions we obtain 
\bea
F_\chi(\widetilde{t}) &=& 
  {1\over \widetilde{t}} \left[1 +
     {1\over 12 \pi} \widetilde{t}^{-1/2} - 
     {1\over 288 \pi^2 \widetilde{t}} 
           \log\left({36\pi^2 {\widetilde{t}}}\right)\right.
\nonumber \\
&& \quad \left.
   - {59\over 2592\pi^2} {1\over \widetilde{t}} +
   {D_3\over \widetilde{t}} + O(\widetilde{t}^{-3/2} 
   \log \widetilde{t})\right].
\eea
\bea
F_\xi(\widetilde{t}) &=& 
  {1\over \widetilde{t}} \left[1 +
     {1\over 12 \pi} \widetilde{t}^{-1/2} - 
     {1\over 288 \pi^2 \widetilde{t}} 
           \log\left({36\pi^2 {\widetilde{t}}}\right)\right.
\nonumber \\
&& \quad \left.
   - {11\over 486\pi^2} {1\over \widetilde{t}} +
   {D_3\over \widetilde{t}} + O(\widetilde{t}^{-3/2} 
   \log \widetilde{t})\right].
\eea

\subsubsection{Asymptotic behavior near the Wilson-Fisher point} 
\label{secA.1.2}

Let us now compute the asymptotic behavior of the crossover functions 
for $\widetilde{t}\to 0$. This requires the determination  of the 
expansion of the various RG
functions in the limit $g\to g^*$. As it has been 
extensively discussed in the literature 
\cite{Parisi_Cargese,Parisi_78_79,Nickel_Cargese,Nickel_91,%
Murray-Nickel_91,Sokal_94,Bagnuls-Bervillier_97,PV_gstar,CCCPV-00,CaPeVi-00},
these functions are singular at $g = g^*$. General arguments predict a 
behavior of the form \cite{Nickel_Cargese,PV_gstar}
\begin{eqnarray} 
W(g) &=& - \omega (g^* - g)\, +
    w_1 (g^* - g)^2 + w_2 (g^* - g)^{\Delta_2/\Delta}
           + w_3 (g^* - g)^{1+1/\Delta} + \cdots
\\ 
\gamma(g) &=& \gamma +
    \gamma_1 (g^* - g) + \gamma_2 (g^* - g)^2 
           + \gamma_3 (g^* - g)^{\Delta_2/\Delta}
           + \gamma_4 (g^* - g)^{1/\Delta} + \cdots
\\ 
   \nu(g) &=& \nu +
   \nu_1 (g^* - g) + \nu_2 (g^* - g)^2 
           + \nu_3 (g^* - g)^{\Delta_2/\Delta}
           + \nu_4 (g^* - g)^{1/\Delta} + \cdots
\end{eqnarray}
This nonanalytic
behavior makes the determination of the corrections extremely 
difficult. For instance, since one expects $\Delta_2/\Delta$ to be 
close to 2 \cite{Riedel}, it is practically impossible
to determine $w_1$ and $w_2$ in the $\beta$-function
since these two terms are essentially degenerate. 
The only subleading coefficients that can be reliably determined are 
$\gamma_1$ and $\nu_1$. Indeed, since $\Delta\approx1/2$
and $\Delta_2/\Delta\approx 2$, the next-to-leading correction
behave as $(g^* - g)^{\approx 2}$, so that 
$\gamma'(g)$ and $\nu'(g)$ should be reasonably smooth 
for $g\to g^*$.

The computation is straightforward and we 
only report the final results. The crossover functions 
can be expanded for $\widetilde{t}\to 0$ as 
\begin{eqnarray}
F_\chi(\widetilde{t}) &=&
  \chi_0 \widetilde{t}^{\, -\gamma} \left[1 + 
   \chi_1 \widetilde{t}^{\Delta} + 
   \chi_2 \widetilde{t} + 
   \chi_3 \widetilde{t}^{\, \Delta_2} + 
   \chi_4 \widetilde{t}^{\, 2 \Delta} + \cdots\right],
\label{B8}
\\
F_\xi(\widetilde{t}) &=&
  \xi_0^2 \widetilde{t}^{-2\nu} \left[1 + 
   \xi_1 \widetilde{t}^{\Delta} + 
   \xi_2 \widetilde{t} + 
   \xi_3 \widetilde{t}^{\Delta_2} + 
   \xi_4 \widetilde{t}^{2 \Delta} + \cdots\right].
\label{B9}
\end{eqnarray}
We obtain for the leading term and the first correction:
\begin{eqnarray}
\chi_0 &=& 
   (6 \pi g^*)^2 \widehat{t}^{\, \gamma}
   \exp\left\{ - \int_0^{g^*} dx\,
       \left[{\gamma(x)\over \nu(x) W(x)} + {2\over x} + 
             {\gamma\over \Delta (g^*-x)}\right]
       \right\},
\\
\xi_0^2 &=& 
   (6 \pi g^*)^2 \widehat{t}^{\, 2\nu }
   \exp\left\{ - 2 \int_0^{g^*} dx\,
       \left[{1\over W(x)} + {1\over x} + {1\over \omega (g^*-x)}\right]
       \right\},
\\ 
\chi_1 &=& - {g^*\gamma_1\over \Delta (1 + \Delta)} 
                 \widehat{t}^{\, -\Delta},
\\ 
\xi_1 &=& - {2 g^*\over \gamma \Delta (1 + \Delta)} \widehat{t}^{\,-\Delta}\,
   \left[ \gamma \nu_1(1 + \Delta) - \nu \gamma_1 \Delta\right],
\end{eqnarray}
where
\be
\widehat{t} =
 {\gamma\over (6 \pi g^*)^2}\,
   \exp\left\{ \int_0^{g^*} dx\,
    \left[{1\over \nu(x)W(x)} + {2\over x} + {1\over \Delta (g^*-x)}\right]
       \right\}.
\ee
In order to estimate these constants we have used 
the seven-loop results of Ref. \cite{Murray-Nickel_91} and 
the resummation technique of Ref. \cite{LeGuillou-ZinnJustin_80}.
Errors due to the resummation have been determined by using the 
general procedure of Ref. \cite{PV_gstar}. In order to compute 
$\chi_0$ and $\xi_0^2$ we have performed resummations 
keeping $g^*$, $\gamma$, $\nu$, and $\Delta$ as free 
parameters (the dependence 
on $\Delta$ cancels in $\chi_0$ and $\xi^2_0$ 
if one uses the explicit expression for 
$\widehat{t}$). We obtain finally
\begin{eqnarray}
\chi_0 &=& 0.4216 \pm 0.0006 - 4 (\gamma - 1.1575) - 0.1 (g^* - 1.395),
\\
\xi_0^2 &=& 0.3565 \pm 0.0002 + 0.4 (\gamma - 1.1575)
             - 0.1 (g^* - 1.395) - 8 (\nu - 0.58758) ,
\\
\widehat{t} &=&
     \left[0.99 \pm 0.004 + 0.9 (\gamma - 1.1575) 
              - 8 (g^* - 1.395) - 9 (\Delta - 0.515)\right]\cdot 10^{-3},
\end{eqnarray}
where the first error is related to the uncertainty in the resummation, while 
the other terms indicate the variation of the estimate with changes 
in the values of the critical exponents and of $g^*$. To obtain the final 
results, we must decide which estimates to use for $\gamma$, $\nu$,
$\Delta$, and 
$g^*$. In principle, we could use the values that have been 
determined from the resummation of the perturbative 
expansion in three dimensions \cite{Guida-ZinnJustin_98},
i.e. $g^* = 1.413 \pm 0.006$, $\gamma = 1.1596 \pm 0.0020$, 
$\nu = 0.5882 \pm 0.0011$, and $\Delta = 0.478 \pm 0.010$. 
However, we believe the Monte Carlo estimates of the critical 
exponents to be more reliable, and thus we have used the 
values reported in Eq. \reff{esponenticritici}. The field-theoretic
estimate of $g^*$ is probably also not reliable since it differs from the 
estimates obtained using different methods. 
Indeed, from the $\epsilon$-expansion one estimates 
$g^* = 1.396\pm 0.020$ \cite{PV-00}, while the extrapolation of 
exact-enumeration series gives $g^* = 1.388\pm 0.005$ \cite{Butera-Comi_98}:
One observes a systematic discrepancy, which we believe to be due 
to the nonanalytic structure of the $\beta$-function 
that is not properly taken into account in the analysis.\footnote{
A different resummation that tries to keep track of the 
nonanalyticities of the $\beta$-function give \cite{Murray-Nickel_91}
$g^*\approx 1.39$.} This problem should also 
appear in our analysis since we use the same resummation technique.
In the reanalysis of Ref. \cite{Caracciolo-etal_97} of the perturbative 
series for the exponent $\gamma$ 
it was shown that the systematic error could be reduced, obtaining 
field-theory estimates in close agreement with the Monte Carlo results,
if one uses $g^* \approx 1.395$. 
Therefore, we use $g^* = 1.395 \pm 0.015$, where the error is such
to include all estimates.
Our final estimates are
\begin{eqnarray}
\chi_0 &= & 0.4216 \pm 0.0029,  \\
\xi_0^2 &=&   0.3565 \pm 0.0016,  \\
\widehat{t} &=& \left(0.99 \pm 0.21\right)\cdot 10^{-3}.
\end{eqnarray}
To compute $\chi_1$ and $\xi_1$, we have analyzed the series
of the derivative of $\gamma(g)$ and $\nu(g)$. We obtained
\begin{eqnarray}
\gamma_1 &=&  - \gamma'(g^*) =\, 
   - 0.1071 \pm 0.0013 + 0.007 (g^* - 1.395), \\
\nu_1 &=&  - \nu'(g^*) = \,   
   - 0.0659 \pm 0.0018 - 0.011 (g^* - 1.395). 
\end{eqnarray}
The ratio $\gamma_1/\nu_1$ has already been computed in 
Ref. \cite{Murray-Nickel_91} finding
\be
{\gamma_1\over \nu_1} =
   1.31 \pm 0.05 - 1.7 (g^* - 1.39) 
\label{ratioNickel}
\ee
that however differs significantly from our result 
\be 
{\gamma_1\over \nu_1}  =
   1.62 \pm 0.05 - 0.4 (g^* - 1.395).
\ee

Using the estimates of the critical exponents reported 
in Eq. \reff{esponenticritici}
and, as before, $g^* = 1.395\pm 0.015$, we obtain
\begin{eqnarray}
\chi_1 &=&   6.8 \pm 0.8, \\
\xi_1 &=&    9.1 \pm 1.1.
\end{eqnarray}
Notice that a significant fraction of the error is due to the 
uncertainty on $\widehat{t}$. The error is largely reduced if we 
consider the ratio $\chi_1/\xi_1$. We obtain
\be
{\chi_1\over \xi_1} =  0.745 \pm 0.034.
\ee
Notice that, if we use the estimate \reff{ratioNickel} we would obtain 
$\chi_1/\xi_1\approx 0.56$. The ratio $\chi_1/\xi_1$ can also be computed
in $\epsilon$-expansion using the $O(\epsilon^2)$ series of 
Ref. \cite{Nicoll-Albright_85}:
\be
{\chi_1\over \xi_1} =  1 - {\epsilon\over8} - 
   \left( {\lambda\over12} + {31\over256}\right)\epsilon^2 + 
    O(\epsilon^3),
\ee
where $\lambda \approx 1.171854$.
We obtain $\chi_1/\xi_1 = 0.85 \pm 0.10$, where the error is purely
indicative because of the shortness of the series.

\subsection{Crossover functions in dimensional regularization without
$\epsilon$-expansion} \label{secA.2}

\subsubsection{General results} \label{secA.2.1}

In this Section  we will study
the critical crossover functions using
the minimal renormalization scheme proposed in 
Refs.~\cite{Dohm,Schloms-Dohm_89,K-S-D}.
We start from the expressions~\cite{Schloms-Dohm_89,K-S-D}
\begin{eqnarray}
F_\chi(\widetilde{t}) &=& \chi^* F(u)^{-1} 
   \, \exp\left[ - \int_{u_i}^u dx\, {\gamma_M(x)\over \nu_M(x) W_M(x)}\right], \\
F_\xi(\widetilde{t}) &=& \left(\xi^*\right)^2
   \, \exp\left[ - 2 \int_{u_i}^u dx\, {1\over W_M(x)}\right], 
\end{eqnarray}
where $\widetilde{t}$ is related to the minimal-subtraction renormalized 
coupling $u$ by
\begin{equation}
\widetilde{t} = 
  - t_0   \int^{u^*}_u dx\, 
 {2 P(x)\over W_M(x)} 
  \exp\left[ \int_{u_i}^x dz\, {1\over \nu_M(z) W_M(z)} \right].
\label{ttilde}
\end{equation}
Here $\gamma_M(x)$, $\nu_M(x)$, and $W_M(x)$ are the standard RG 
functions computed in
dimensional regularization, $P(x)$  and $F(x)$ are
functions defined in Refs. \cite{Schloms-Dohm_89,K-S-D} 
that will be explicitly given below, 
and $u^*$ the critical value of $u$ defined by
$W_M(u^*) = 0$. The constant $\chi^*$, $\xi^*$, $t_0$, and $u_i$ are 
obtained by requiring that, for $\widetilde{t}\to 0$, 
$F_\chi(\widetilde{t})$ and $F_\xi(\widetilde{t})$ behave as in
Eq. \reff{fchiximf}.
The RG functions have been computed to five-loop order
in Refs.\cite{C-G-L-T,K-N-S-C-L}.
Explicitly we have\footnote{We normalize $u$ as in 
Ref.~\cite{ZinnJustin_book}.
Our coupling $u$ differs from the coupling  used in 
Refs. \cite{Schloms-Dohm_89,K-S-D}
by a factor of $4!$ and from
the definition used in Refs. \cite{C-G-L-T,K-N-S-C-L}
by a factor of 2.}
\begin{eqnarray}
W_M(u) &=& W_M(u,1), \\
\nu_M(u) &=& {1\over 2 + \eta_2(u)} ,\\
\gamma_M(u) &=& \left[ 2 - \eta_3(u) \right] \nu_M(u) ,
\end{eqnarray}
where
\begin{eqnarray}
&& W_M(u,\epsilon) =
  -\epsilon u + \case{4}{3} u^2 - \case{7}{6}u^3 + 
   \case{2960 + 2112 \zeta (3)}{1728}u^4 + 
\case{-196648 + 2816\pi/5 - 
         223872\zeta (3) - 357120 \zeta (5)}{62208} u^5
     + \nonumber \\
&& \case{ 13177344 - 67584 \pi^4 - 
         317440 \pi^6/21 + 21029376\zeta (3) + 
         2506752\zeta (3)^2 + 42261504\zeta (5) + 
         59383296\zeta (7)}{1990656} u^6 + O(u^7), 
\end{eqnarray} 
\begin{eqnarray}
&&  \eta_2(u) = 
  -\case{1}{3}u + \case{5}{36}u^2 
- \case{37}{144} u^3 - 
   \case{-31060 - 352\pi^4/5 - 
         3264\zeta (3)}{62208} u^4  -
\nonumber \\
&&   \case{3166528 + 42688\pi^4/5 + 
         39680\pi^6/21 + 1528704\zeta (3) - 
         446976\zeta (3)^2 + 55296\zeta (5)}{2985984} u^5 + O(u^6),
\end{eqnarray}
\begin{equation} 
   \eta_3(u) =   \case{1}{36}u^2 - \case{1}{108} u^3  + \case{125}{5184} u^4 + 
   \case{ -77056 - 1408\pi^4/5 + 
         8832\zeta (3)}{1492992} u^5 + O(u^6).
\end{equation}
To compute the functions $P(u)$ and $F(u)$ let us first recall the relation 
between the bare coupling $u_0$  and the renormalized coupling $u$ 
\begin{eqnarray}
&&u_0  = \mu A_3^{-1} u Z_u(u) Z_\varphi(u)^{-2},\\
&&Z_u(u) Z_\varphi(u)^{-2} = 
   \exp \left\{ - \int_0^u d u' 
        \left[{1\over W_M(u')} + {2\over u'} \right] \right\}.
\label{u0}
\end{eqnarray}
Here $\mu$ is the renormalization scale and $A_d$ is a constant that depends on the specific 
renormalization scheme. Of course physical results should not depend on it.
This fact can be easily verified
noticing that $A_d$ can be absorbed in the definition of $\mu$, and that,
by construction, all physical quantities are independent of $\mu$.
However, different choices of $A_d$
give rise to different perturbative series providing different results at the 
intermediate stages of the calculation. This freedom may be used
as a further check of the real uncertainty of the final results.
In Refs. \cite{Schloms-Dohm_89,K-S-D} the authors use 
\begin{eqnarray}
A_d &=&  S_d \Gamma(3-d/2) \Gamma(d/2-1),\label{ad} \\
S_d &=& 
   {2\over (4\pi)^{d/2} \Gamma(d /2)}, \nonumber
\end{eqnarray}
a choice 
that makes the one-loop corrections vanish in many observables. However,
in order to understand the size of the systematic errors, we will also
use $A_d = S_d$ and $A_d = 4 S_d$.
The functions $F(u)$ and $P(u)$ are obtained from
\begin{eqnarray}
&& 
F(u) \equiv Z_\varphi(u) F_0( A_3^{-1} 
    u Z_u(u)Z_\varphi(u)^{-2}),
\\
&&
Z_\varphi(u)  = \exp\left[ \int_0^u
du' {\eta_3(u')\over W_M(u')}\right] , 
\end{eqnarray}
and
\begin{eqnarray}
&& 
P(u) \equiv 
Z_{\varphi^2}(u)^{-1} P_0(A_3^{-1} u Z_u(u)Z_\varphi(u)^{-2}),\\
&&
Z_{\varphi^2}(u) = \exp\left[ \int_0^u
du' {\eta_2(u')\over W_M(u')}\right] ,
\end{eqnarray}
where $F_0(x)$ and $P_0(x)$ can be derived from 
the five-loop results of Refs. \cite{Bagnuls-Bervillier_85,K-S-D}:
\begin{eqnarray}
F_0(x) &=& 1 + \case{1}{7776\pi^2} x^2
- 8.83291\cdot {{10}^{-7}}\,{x^3} +
   6.17241\cdot {{10}^{-8}}\,{x^4} 
\nonumber \\ 
&& - 4.73993\cdot {{10}^{-9}}\,{x^5} + O(x^6),\\
P_0(x)&=&  1 + \case{1}{24\pi} x -\case{1}{288\pi^2} x^2
+ 1.81785\cdot 10^{-5}\,\,{x^3} -
   1.24518\cdot{{10}^{-6}}\,{x^4} 
\nonumber \\
&& + 1.01097\cdot{{10}^{-7}}\,{x^5} + O(x^6).
\end{eqnarray}
When $A_d$ is given by Eq.~(\ref{ad}) we have
\begin{eqnarray}
F(u)&=&  1 - \case{23}{1944} u^2
- 5.52510\,10^{-3}\,{u^3} - 4.01633\,10^{-3}\,{u^4} - 1.92954\,10^{-3}\,{u^5}
+ O(u^6),  \nonumber \\ 
&& \\
P(u) & = &
  1 - \case{1}{6} u + \case{1}{72}u^2
- 0.0318279\,{u^3} + 0.0326664\,{u^4} - 0.0644065\,{u^5}
+ O(u^6). \nonumber \\ 
&& 
\end{eqnarray}

\subsubsection{Asymptotic behavior near the Wilson-Fisher point} 
\label{secA.2.2}

In order to compute the asymptotic behavior of the crossover functions 
for $\widetilde{t}\to 0$, we need the expansion of the various 
RG functions in the limit $u\to u^*$. It has been argued
\cite{Schafer_94} that the scheme we are presenting 
has an important advantage over the 
approach described in App. \ref{secA.1}: The RG
functions are expected to be analytic at the critical point. 
The reason is that the RG functions are essentially
dimension-independent, while $u^*$ depends on $\epsilon$ being the 
solution of $W_M(u^*,\epsilon)=0$. Notice, however, that this argument 
does not exclude the presence of singular terms for $u=0$ since this 
scheme is essentially four-dimensional. Answering this question is in 
any case nontrivial since it requires a nonperturbative definition
of the RG functions. In the following we will 
assume the following analytic expansions:
\begin{eqnarray}
W_M(u) &=& - \omega (u^* - u)\, +
    w_1 (u^* - u)^2 + \ldots \\
\gamma_M(u) &=& \gamma +
    \gamma_1 (u^* - u) + \ldots \\
\nu_M(u) &=& \nu +
   \nu_1 (u^* - u) + \ldots \\
F(u) &=& f^* +
   f_1 (u^* - u) + \ldots \\
P(u) &=& p^* +
   p_1 (u^* - u) + \ldots 
\end{eqnarray}
Expanding the crossover functions 
for $\widetilde{t}\to 0$ as in Eqs. \reff{B8} and \reff{B9},
we obtain for the leading term and the first correction:
\begin{eqnarray}
&&
\chi_0  = 
   {(\kappa u^*)^2\over f^*}  \widehat{t}^{\, \gamma}
   \exp\left\{ - \int_0^{u^*} dx\,
       \left[ {\gamma_M(x)\over \nu_M(x) W_M(x)} + {2\over x} + 
             {\gamma\over \Delta (u^*-x)}\right]
       \right\},
\\
&&
\xi_0^2  = 
   (\kappa u^*)^2 \widehat{t}^{\, 2\nu }
   \exp\left\{ - 2 \int_0^{u^*} dx\,
       \left[ {1\over W_M(x)} + {2\over x} + {1\over \omega (u^*-x)}\right]
       \right\},
\\
&& 
\chi_1 =  - u^* \left[ {f_1\over f^*}  + {\gamma_1\over \Delta}
-{\gamma\over 1 + \Delta} \left({\nu_1\over \nu} + 
      {p_1\over p^*}\right)\right]   
                 \widehat{t}^{\, -\Delta},
\\
&& 
\xi_1 = 
  u^* {2 \nu\over 1 + \Delta} 
     \left(  {p_1\over p^*} - {\nu_1\over \Delta \nu} \right)   
                 \widehat{t}^{\, -\Delta},
\end{eqnarray}
where
\begin{equation}
\widehat{t} =
 {2p^* \nu \over (\kappa  u^*)^2}\,
   \exp\left\{ \int_0^{u^*} dx\,
    \left[ {1\over \nu_M(x) W_M(x)} + {1\over x} + 
           {1\over \Delta (u^*-x)}\right]
       \right\}.
\end{equation}
The normalization $\kappa$ is related to the choice of $A_d$ in 
Eq.~(\ref{u0}) by
\begin{eqnarray}
\kappa &=& 4\pi (1 + 2 q_3), \\ 
q_d &=& { S_d \Gamma(3-d/2)\Gamma(d/2-1)- A_d \over 2 (4-d) A_d}\nonumber
\end{eqnarray} 
The estimate of critical quantities requires a resummation of the perturbative
series.  For this purpose we will use 
the large-order behavior of the coefficients~\cite{Lipatov,B-L-Z} given by
\begin{equation}
c_k \sim k! \left(- \case{1}{2} \right)^k k^b \left[ 1 + O(1/k)\right],
\end{equation}
and the numerical method of Refs. \cite{LeGuillou-ZinnJustin_80,PV_gstar}.

We have first of all determined $u^*$. We obtain $u^* = 1.1\pm0.1$. 
Although this result is consistent with the estimate of Ref. 
\cite{Schloms-Dohm_89}, $u^* = 1.092 \pm 0.012$, the error bar is much larger.
However, using our algorithm \cite{PV_gstar}, we have been unable
to understand how the error can be so small. On the other hand,
as we shall see later by comparing 
our results with the estimates of the previous Section and by checking 
their independence on $A_d$, our error bars look reasonable 
and at most overestimated by a factor of two. In the following we report 
various estimates keeping $u^*$ as a free variable. We have
\begin{eqnarray}
\gamma_1 &=& - 0.140 \pm 0.010 - 0.01 (u^* - 1.1), \\
\nu_1 &=& - 0.089 \pm 0.010 - 0.02 (u^* - 1.1), \\
f^* &=& \cases{0.991 \pm 0.004 - 0.02 (u^* - 1.1), & \cr
               0.987 \pm 0.006 - 0.02 (u^* - 1.1), & \cr
               0.983 \pm 0.006 - 0.02 (u^* - 1.1), & } \\
p^* &=& \cases{0.932 \pm 0.006 - 0.05 (u^* - 1.1), & \cr
               0.825 \pm 0.015 - 0.15 (u^* - 1.1), & \cr
               0.670 \pm 0.040 - 0.30 (u^* - 1.1),  & } \\
f_1 &=& \cases{0.026 \pm 0.007 + 0.03 (u^* - 1.1), & \cr
               0.039 \pm 0.012 + 0.04 (u^* - 1.1), & \cr
               0.051 \pm 0.018 + 0.06 (u^* - 1.1), & } \\
p_1 &=& \cases{0.079 \pm 0.015 + 0.03 (u^* - 1.1), & \cr
               0.182 \pm 0.012 + 0.02 (u^* - 1.1), & \cr
               0.399 \pm 0.030 + 0.13 (u^* - 1.1). & }
\end{eqnarray}
For $f^*$, $p^*$, $f_1$, and $p_1$ we report three estimates 
corresponding to $A_3 = S_3,1/4\pi,4S_3$ respectively. Notice that in most 
cases the uncertainty on $u^*$ is negligible compared to the resummation errors.
We obtain finally
\begin{eqnarray}
\chi_0 &=& 0.420(15),0.422(19),0.447(42), \\
\xi_0^2 &=& 0.358(15),0.363(19),0.395(38), \\
\chi_1  &=& 6.7(4.2),6.5(4.1),5.4(3.4), \\
\xi_1   &=& 10.1(6.3),9.2(5.7),6.7(4.2),
\end{eqnarray}
where the three different estimates correspond to 
$A_3 = S_3,1/4\pi,4S_3$ respectively. As before, a more precise estimate is
obtained if one considers $\chi_1/\xi_1$. We obtain
\be
{\chi_1\over\xi_1} = 
    0.66(13),0.71(10),0.81(7).
\ee
As expected, these results 
are independent of the value of $A_3$ within error bars.
Notice that the difference among the estimates of the same quantity is 
of the same order of the error bars, thereby confirming the correctness
of our error estimates. The final results are also in good 
agrement with, although less precise than, the results presented in the 
previous Section. Notice that the estimate of $\chi_1/\xi_1$ obtained 
here is compatible with the estimate obtained in Sec. \ref{secA.1.2}.
Instead, the result of Ref. \cite{Murray-Nickel_91} is somewhat too small.

\subsection{Polymer critical crossover functions} \label{secA.3}

Let us now compute the critical crossover functions in terms of 
$\widetilde{n}$. We will only consider the approach described in 
Sect. \ref{secA.1}, since it appears to be the most precise one.
From Eq. \reff{fchi-gc} we have
\be
G_c(\widetilde{n})\, =\, 
   \int^{c+i\infty}_{c-i\infty} {d\widetilde{t}\over 2\pi i}\, 
    e^{\widetilde{n}\, \widetilde{t}}\, F_\chi(\widetilde{t}).
\ee
Changing variables from $\widetilde{t}$ to $g$ we obtain
\be
G_c(\widetilde{n})\, =\, \chi^* t_0 
 \int_C {dg\over 2 \pi i}\, e^{\widetilde{n}\, \widetilde{t}(g)}\,
   {\gamma(g)\over \nu(g) W(g)}\, 
   \exp\left[ \int_{y_0}^g dz\, {1-\gamma(z)\over \nu(z) W(z)}\right],
\ee
where $C$ may be taken as a circle of the form 
$g = R(1 + e^{-i\phi})$, $-\pi\le \phi \le \pi$, with $R$ fixed
satisfying $0<R<g^*/2$.

To compare with the results of Ref. \cite{Muthukumar-Nickel_87} we introduce 
their notations
\begin{eqnarray}
&& \beta(g)\, = \, {2\nu(g) W(g)\over \gamma(g)}, \\
&& j(g)\, =\, 2 \left(1 - {1\over \gamma(g)}\right),
\\
&& J(g)\, =\, \left(1 - {g\over g^*}\right)^{(1-\gamma)/\Delta}\, 
   \exp\left\{ -\int_0^g dx\left[{j(x)\over \beta(x)} + 
    {\gamma - 1\over \Delta (g^*-x)}\right]\right\} ,
\\
&& A(g)\, =\, \exp\left\{\int_0^g dx\left[{2-j(x)\over \beta(x)} + {2\over x}
    + {1\over \Delta (g^*-x)}\right]\right\} ,
\\
&& E_0(g)\, =\, \int_{g^*}^g dx\, 
          \left(1 - {x\over g^*}\right)^{1/\Delta}\, 
          {A(x)\over x^2 \beta(x)},
\end{eqnarray}
where $\gamma$ and $\Delta$ are the standard critical exponents. Then, we obtain
\be
G_c(\widetilde{n})\, =\, 2 \, \int_C
        {dg\over 2 \pi i} {J(g)\over \beta(g)} \,
        \exp\left[ {\widetilde{n} \over 18 \pi^2}\, 
                 E_0(g)\right].
\ee
In order to have the same definitions of Refs. 
\cite{Muthukumar-Nickel_84,Muthukumar-Nickel_87}, let us also introduce
\be 
z = \, {\sqrt{\widetilde{n}}\over 24 \pi^{3/2}}.
\ee
We obtain finally
\be
G_c(z)\, =\, \int_C {dg\over \pi i}\, {J(g)\over \beta(g)}\, 
    e^{32 \pi z^2 E_0(g)}.
\label{Gcfinale}
\ee
Expanding the previous expression for $z\to 0$, we obtain 
the perturbative expansion \reff{Gcpert}.

The computation of $G_E(z)$ is analogous. Starting from 
\be
G_E(\widetilde{n})\, =\, {1\over G_c (\widetilde{n})}
  \int^{c+i\infty}_{c-i\infty} {d\widetilde{t}\over 2\pi i}\,
    e^{\widetilde{n}\, \widetilde{t}}\, F_\chi(\widetilde{t})
                         F_\xi (\widetilde{t}),
\ee
we obtain for the swelling factor \cite{Muthukumar-Nickel_87}
$S_E(z) = G_E(z)/(6 \widetilde{n})$
\be
S_E(z)\, =\, {1\over 16 \pi z^2 G_c(z)} \, \int_C
   {dg\over \pi i}\, {J(g)K(g)\over \beta(g)E(g)}\, e^{32 \pi z^2 E_0(g)},
\label{hEfinale}
\ee
where
\begin{eqnarray}
&& k(g) \, =\, 2\left( {2\nu(g)\over \gamma(g)} - 1\right),
\\
&& K(g) \, =\, 
   \left(1 - {g\over g^*}\right)^{(\gamma-2\nu)/\Delta}\,
   \exp\left\{ -\int_0^g dx\left[{k(x)\over \beta(x)} +
    {\gamma - 2\nu\over \Delta (g^*-x)}\right]\right\} ,
\\
&& E(g) \, =\, 
     {1\over g^2} 
   \left(1 - {g\over g^*}\right)^{\gamma/\Delta}\,
   \exp\left\{ 2\int_0^g dx\left[{1\over \beta(x)} + {1\over x} +
    {\gamma\over 2\Delta (g^*-x)}\right]\right\} .
\end{eqnarray}
From the previous expressions we can compute the asymptotic behavior
of $G_c(z)$ and $S_E(z)$ for $z\to\infty$. We have
\begin{eqnarray}
G_c(z) &=& g_{c0} z^{2(\gamma - 1)} \left(
    1 + g_{c1} z^{-2\Delta} + g_{c2} z^{-2} + g_{c3} z^{-2 \Delta_2} +
    \ldots \right), 
\label{hcztoinfty}\\
S_E(z) &=& s_{E0} z^{4 \nu - 2} \left(
    1 + s_{E1} z^{-2\Delta} + s_{E2} z^{-2} + s_{E3} z^{-2 \Delta_2} +
    \ldots \right). 
\label{hEztoinfty}
\end{eqnarray}
These expansions can be related to the expansions of $F_\chi(\widetilde{t})$
and $F_\xi(\widetilde{t})$ for $\widetilde{t}\to 0$. Using the 
results of App. \ref{secA.1.2}, we have
\begin{eqnarray}
&& 
g_{c0} \, =\, {\chi_0\over \Gamma(\gamma)} 
     \left(24 \pi^{3/2}\right)^{2(\gamma-1)}\, =\, 
     2.117 \pm 0.020,
\label{hc0numerico}
\\
&& 
s_{E0} \, =\, {\xi_0^2 \Gamma(\gamma)\over \Gamma(\gamma + 2 \nu)}
     \left(24 \pi^{3/2}\right)^{2(2 \nu-1)} \, =\,
     1.549 \pm 0.007,
\\
&& 
g_{c1} \, =\, {\chi_1 \Gamma(\gamma)\over \Gamma(\gamma-\Delta)}
        \left(24 \pi^{3/2}\right)^{-2\Delta} \, =\,
     0.029 \pm 0.005,
\label{hc1numerico}
\\
&& 
s_{E1} \, =\, \left[
           {(\chi_1 + \xi_1) \Gamma(\gamma + 2 \nu)\over 
               \Gamma(\gamma + 2 \nu - \Delta)} - 
            {\chi_1 \Gamma(\gamma)\over \Gamma(\gamma-\Delta)}\right]\,
        \left(24 \pi^{3/2}\right)^{-2\Delta} \, =\, 0.101 \pm 0.016.
\end{eqnarray}
If we consider the ratio $g_{c1}/s_{E1}$ the error is largely reduced 
and we have 
\be
{g_{c1}\over s_{E1}}\, =\, 0.288 \pm 0.016.
\label{hc1suhE1numerico}
\ee
The constants $s_{E0}$ and $s_{E1}$ have already been computed in Ref.
\cite{Muthukumar-Nickel_87}, finding 
$s_{E0} \approx 1.5310$ and $s_{E1} \approx 0.1204$, in reasonable agreement
with our results. The constants $s_{E0}$ and $s_{E1}$ have also been
determined by a Monte Carlo simulation of the Domb-Joyce model
\cite{Belohorec-Nickel_97}, obtaining
\bea
s_{E0} &=& \lim_{\omega\to0} B_R(\omega) \approx 1.54654  \\
s_{E1} &=& \lim_{\omega\to0} b_R(\omega) \approx 0.11498  
\eea
where $B_R(\omega)$ and $b_R(\omega)$ are defined in 
Ref. \cite{Belohorec-Nickel_97}.

We have computed the functions $G_c(z)$ and $S_E(z)$ using the 
numerical technique presented in Ref. \cite{Muthukumar-Nickel_87}. 
In the resummation
we have used the seven-loop results of Ref. \cite{Murray-Nickel_91} that
allow the extension of the series expansions of $j(g)$ and $k(g)$
by one order. If
\be
j(g)=\, \sum_n j_n g^n,\qquad \qquad
k(g)=\, \sum_n k_n g^n,
\ee
we derive from Ref. \cite{Murray-Nickel_91}
\be
j_7 = 0.0996888, \qquad \qquad k_7 = - 0.00190671.
\ee
The results for $G_c(z)$ are well fitted by
\be
G_c(z) =
         (1 + 50.79365 z + 508.5428 z^2 +
              5929.475 z^3 + 10937.03 z^4)^{0.07875}.
\ee
For $z\to 0$ this expression gives $G_c(z) \approx 1 + 4 z$, in agreement with
the perturbative expansion \reff{Gcpert}, while for $z\to\infty$ we have
\be
G_c(z) = \, 2.08\, z^{2 \gamma - 2}\, (1 + 0.089 z^{-1} + O(z^{-2}) ).
\ee
This expansion is in reasonable agreement with  Eq. \reff{hcztoinfty}, 
keeping into account that $\Delta \approx 1/2$. However, while the leading term 
is close to the estimate \reff{hc0numerico}, the correction differs
significantly from Eq. \reff{hc1numerico}. A simpler expression, that is however
more accurate in the Wilson-Fisher region, is 
\be
G_c(z) = (1 + 38.0952 z + 276.844 z^2 + 1073.17 z^3)^{0.105}.
\label{GcresumApp}
\ee
For $z\to\infty$ it behaves as 
\be
G_c(z) \approx \, 2.08\, z^{2 \gamma - 2}\, (1 + 0.027 z^{-1}),
\ee
in agreement with the asymptotic expansion
\reff{hcztoinfty} and the numerical values
\reff{hc0numerico}, \reff{hc1numerico}.

For $S_E(z)$ we find that the expression reported
in Ref. \cite{Belohorec-Nickel_97}
\be
S_E(z) = \,
(1 + 7.6118 z + 12.05135 z^2)^{0.175166}\; ,
\label{SE-BN}
\ee
fits the data extremely well and correctly reproduces the asymptotic behavior 
for $z\to 0$ and $z\to \infty$.

We mention that a different interpolation appears in Ref. \cite{Schafer_94},
based on the five-loop computations of Schloms and Dohm
\cite{Schloms-Dohm_89}. Setting $z = 6.95 \widetilde{z}$ in the 
formulae of Ref. \cite{Schafer_94} in order to reproduce the correct 
behavior for $z\to 0$, we obtain
\be
S_E(z) = \,
   (1 + 7.4074 z + 10.913 z^2)^{0.18},
\ee
in good agreement with Eq. \reff{SE-BN}.
Finally, from the results reported in 
Ref. \cite{Grassberger-etal_97}, we can estimate the ratio 
$g_{c1}/s_{E1}$. We find $g_{c1}/s_{E1} \approx 0.26$, 
in reasonable agreement with Eq. \reff{hc1suhE1numerico}. 
Let us finally notice that if we use the estimate 
\reff{ratioNickel} we would obtain $g_{c1}/s_{E1} \approx 0.23$.

\clearpage

\begin{table}
\squeezetable
\caption{\label{tablerawdata}
Monte Carlo results. Here $\beta_{\rm mf} = (V_R - 1)^{-1}$.
 }
\begin{tabular}{rrcc}
$\rho$ & $n$ & $\log E^2_n$  & $\log(c_n \beta_{\rm mf}^{-n})$ \\
\hline
2 &     20 &      4.094106(9)  &    $-$1.570800(12)  \\
  &  30 &      4.556891(21)  &    $-$2.52151(4) \\
  &  40 &      4.886886(17)  &    $-$3.48884(3)  \\
  &  80 &      5.68682(3)  &    $-$7.42459(6)  \\
  & 160 &      6.49153(6)  &   $-$15.39944(14)  \\
  &  320 &      7.29966(12)  &   $-$31.45439(30)  \\
  &  640 &      8.10973(24)  &   $-$63.67150(64)  \\
  & 1280 &      8.92194(54)  &  $-$128.2124(16)  \\
 \hline
3 &   120 &      6.50585(3)  &    $-$5.20178(6)  \\
  &  160 &      6.82940(15)  &    $-$7.06537(30)  \\
  &  240 &      7.28750(5)  &   $-$10.81482(13)  \\
  & 320 &      7.61466(26)  &   $-$14.58095(62)  \\
  &  480 &      8.07689(10)  &   $-$22.13565(27)  \\
  &  960 &      8.87276(19)  &   $-$44.87624(56)  \\
  & 1920 &      9.67317(37)  &   $-$90.4588(12)  \\
  & 3840 &     10.47830(81)  &  $-$181.7297(31)  \\
\hline
4 &   1200 &      9.39403(32)  &    $-$30.4356(12)  \\
  & 1280 &      9.46739(15)  &    $-$32.4988(5)  \\
  & 1920 &      9.92722(12)  &    $-$49.0221(4)  \\
  & 2560 &     10.25489(28)  &    $-$65.5637(10)  \\
  & 3840 &     10.71901(22)  &    $-$98.6670(8)  \\
  & 5120 &     11.04855(54)  &   $-$131.7917(21)  \\
  & 7680 &     11.51585(45)  &   $-$198.0564(17)  \\
  & 10240 &     11.8470(11)  &   $-$264.3489(44)  \\
  & 15360 &     12.31571(93)  &   $-$396.9388(36)  \\
  & 30720 &     13.1234(22)  &   $-$794.8075(75)  \\
 \hline
5 &    600 &      8.88068(10)  &     $-$8.81049(25)  \\
  &   1040 &      9.484413(43)  &    $-$15.51334(13)  \\
  & 1200 &      9.64205(17)  &    $-$17.95699(50)  \\
  & 2080 &     10.25375(8)  &    $-$31.42686(26)  \\
  & 2400 &     10.41268(28)  &    $-$36.3313(10)  \\
  & 4160 &     11.03218(14)  &    $-$63.3411(6)  \\
  & 8320 &     11.81856(25)  &   $-$127.2624(11)  \\
  & 16640 &     12.61152(54)  &   $-$255.2014(24)  \\
  & 33280 &     13.4103(11)  &   $-$511.1808(48)  \\
  & 66560 &     14.2166(31)  &  $-$1023.251(12)  \\
\hline
6 &   500 &      8.93513(4)  &    $-$4.61781(8)  \\
  &   800 &      9.43841(8)  &    $-$7.52107(20)  \\
  & 1000 &      9.67871(6)  &    $-$9.46419(16)  \\
  & 1600 &     10.18886(14)  &   $-$15.31451(40)  \\
  & 2000 &     10.43242(10)  &   $-$19.22308(33)  \\
  & 3200 &     10.94927(23)  &   $-$30.97376(82)  \\
  & 4000 &     11.19612(18)  &   $-$38.81610(66)  \\
  & 8000 &     11.96920(31)  &   $-$78.0846(13)  \\
  & 16000 &     12.75136(54)  &  $-$156.7110(27)  \\
  & 32000 &     13.5400(11)  &  $-$314.0585(56)  \\
\end{tabular}
\end{table}

\begin{table}
\squeezetable
\caption{\label{tablerawdata2}
Monte Carlo results. Here $\beta_{\rm mf} = (V_R - 1)^{-1}$.
 }
\begin{tabular}{rrcc}
$\rho$ & $n$ & $\log E^2_n$  & $\log(c_n \beta_{\rm mf}^{-n})$ \\
\hline
7 &    100 &      7.500434(17)  &    $-$0.543304(13)  \\
  &  200 &      8.214646(25)  &    $-$1.164668(26)  \\
  &  400 &      8.93561(4)  &    $-$2.44010(5)  \\
  &  800 &      9.66443(6)  &    $-$5.03235(11)  \\
  & 1600 &     10.40211(9)  &   $-$10.26761(23)  \\
  & 3200 &     11.14984(16)  &   $-$20.79866(51)  \\
  & 6400 &     11.90757(25)  &   $-$41.9304(10)  \\
  & 12800 &     12.67547(43)  &   $-$84.2725(21)  \\
  & 25600 &     13.45153(68)  &  $-$169.0420(42)  \\
  & 51200 &     14.2366(14)  &  $-$338.6733(87)  \\
\hline
8 &    750 &      9.80702(3)  &    $-$3.29563(5)  \\
  & 1500 &     10.53428(5)  &    $-$6.74019(10)  \\
  & 3000 &     11.27031(7)  &   $-$13.67826(20)  \\
  & 6000 &     12.01591(12)  &   $-$27.61282(42)  \\
  & 12000 &     12.77156(19)  &   $-$55.55008(85)  \\
  & 24000 &     13.53685(34)  &  $-$111.5009(17)  \\
  & 48000 &     14.31143(59)  &  $-$223.4866(35)  \\
 \hline
9 &   1040 &      10.34095(12)  &     $-$3.35462(21)  \\
  & 2080 &      11.06434(17)  &     $-$6.8409(4)  \\
  & 4160 &      11.79586(27)  &    $-$13.8574(9)  \\
  & 8320 &      12.5369(4)  &    $-$27.9440(18)  \\
  & 16640 &      13.2868(6)  &    $-$56.1807(36)  \\
  & 33280 &      14.0491(10)  &   $-$112.7270(71)  \\
  & 66560 &      14.8188(20)  &   $-$225.901(14)  \\
\hline
10  & 1500 &     10.89765(8)  &    $-$3.66081(13)  \\
   & 2000 &     11.19598(5)  &    $-$4.91804(11)  \\
  & 3000 &     11.61839(11)  &    $-$7.44241(26)  \\
  & 4000 &     11.91986(8)  &    $-$9.97288(22)  \\
  & 6000 &     12.34684(17)  &   $-$15.04661(52)  \\
  & 8000 &     12.65209(13)  &   $-$20.12732(44)  \\
 & 16000 &     13.39353(20)  &   $-$40.49032(91)  \\
 & 32000 &     14.14511(32)  &   $-$81.2802(18)  \\
 & 64000 &     14.90648(72)  &  $-$162.9324(51)  \\
\hline
12 &   1500 &     11.21510(8)  &     $-$2.19646(10)  \\
  & 2000 &     11.50974(10)  &     $-$2.95223(16)  \\
  & 3000 &     11.92649(14)  &     $-$4.47024(21)  \\
  & 4000 &     12.22324(16)  &     $-$5.9926(4)  \\
  & 6000 &     12.64352(20)  &     $-$9.0463(4)  \\
  & 8000 &     12.94315(25)  &    $-$12.1053(8)  \\
  & 12000 &     13.36771(30)  &    $-$18.2350(9)  \\
  & 16000 &     13.67113(37)  &    $-$24.3704(16)  \\
  & 32000 &     14.40765(57)  &    $-$48.9504(31)  \\
\end{tabular}
\end{table}

\begin{table}
\caption{\label{tablebetac}
Determination of $\beta_{c,R}$ for several values of $\rho$. 
      The reported results are obtained by fitting the numerical data 
      with $n > n_{\rm min}$. 
Fit (a): $V_R \beta_{{\rm eff},R} (n) = V_R \beta_{c,R} + a/{n} + 
       b R^3/{n}^{1.5}$.
Fit (b): $V_R \beta_{{\rm eff},R} (n) = V_R \beta_{c,R} + a/{n} +
       b R^3/{n}^{1.5} + c R^6/{n}^2$.
}
\begin{tabular}{rrllll}
$\rho$ & $n_{\rm min}$ & 
\multicolumn{1}{c}{$V_R \beta_{c,R}$} & 
\multicolumn{1}{c}{$a$} &
\multicolumn{1}{c}{$b$} &
\multicolumn{1}{c}{$c$} \\
\hline
\multicolumn{6}{c}{Fit (a)} \\
\hline
2 &  40 & 1.152388(1) & 0.0725(4) & 0.1044(16) & \\
3 &  80 & 1.0656569(3) & 0.0142(3) & 0.0360(5) & \\
4 &  150 & 1.0342660(1)& $-$0.0040(3)& 0.0174(3)& \\
5 &  260 & 1.0199238(1)& $-$0.0111(1)& 0.0100(1)& \\
6 &  400 & 1.0125749(1)& $-$0.0138(4) & 0.0066(2)& \\
7 &  800 & 1.00840580(4) & $-$0.0164(3) & 0.0053(1)& \\
8 & 1500 & 1.00588845(3) & $-$0.0174(4) & 0.0044(1)& \\
9 & 1040 & 1.00427552(9) & $-$0.0140(10) & 0.0026(2)& \\
10& 3000 & 1.00320022(5) & $-$0.0167(13) & 0.0031(3)& \\
12& 2000 & 1.00192307(7) & $-$0.0080(11) & 0.0010(1)& \\
\hline
\multicolumn{6}{c}{Fit (b)} \\
\hline
2 &  10 & 1.152388(1) & 0.0711(1) & 0.1211(3) & $-$0.01139(6)\\
3 &  15 & 1.0656574(3) & 0.0121(1) & 0.0442(1) & $-$0.00182(1)\\
4 &  40 & 1.0342661(1)& $-$0.0057(1)& 0.0217(1) & $-$0.000589(6)\\
5 &  70 & 1.0199239(1)& $-$0.0130(1)& 0.0132(1) & $-$0.000274(3)\\
6 &  100 & 1.0125750(1)& $-$0.0158(4) & 0.0090(2) & $-$0.000152(6)\\
7 &  140 & 1.00840566(10) & $-$0.0159(4) & 0.0062(1) & $-$0.000081(3)\\
8 &  300 & 1.00588839(5) & $-$0.0176(5) & 0.0055(2) & $-$0.000071(3)\\
9 &  250 & 1.00427551(10) &$-$0.0153(10) & 0.0037(2) & $-$0.000037(3)\\
10&  450 & 1.00320011(4) & $-$0.0143(5) & 0.0030(1) & $-$0.000028(1)\\
12&  400 & 1.00192275(6) & $-$0.0028(5) & 0.0006(1) & $-$0.000003(1)\\
\end{tabular}
\end{table}

\clearpage


\begin{figure}
\vspace*{-1cm} \hspace*{-0cm}
\centerline{\psfig{width=10truecm,angle=-90,file=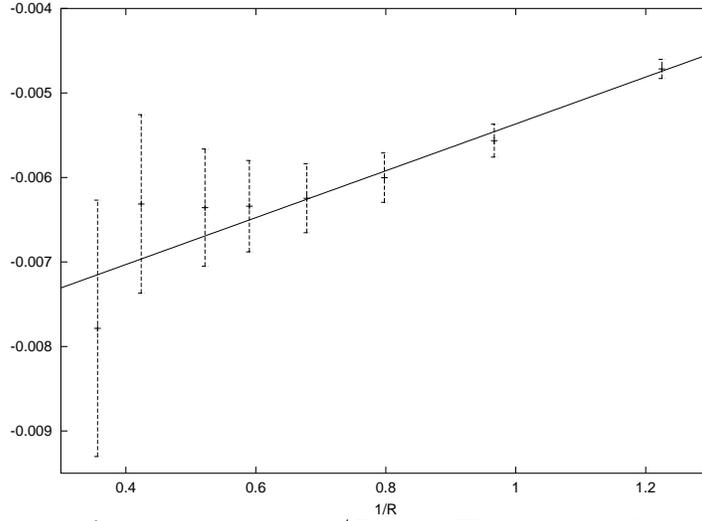}}
\caption{Estimates of $\tau_{2,\rm eff}$ versus $1/R$. 
The reported points correspond to $\rho=3,4,5,6,7,8,10,12$.
The line is the best fit: $\tau_{2,\rm eff} = -0.00814+0.00277/R$.
The errors on the data take into account the error on $\beta_{c,R}$ 
and on the constant $\tau_1$.
}
\label{tau2}        
\end{figure}

\begin{figure}
\begin{tabular}{c}
\centerline{\psfig{width=9truecm,angle=-90,file=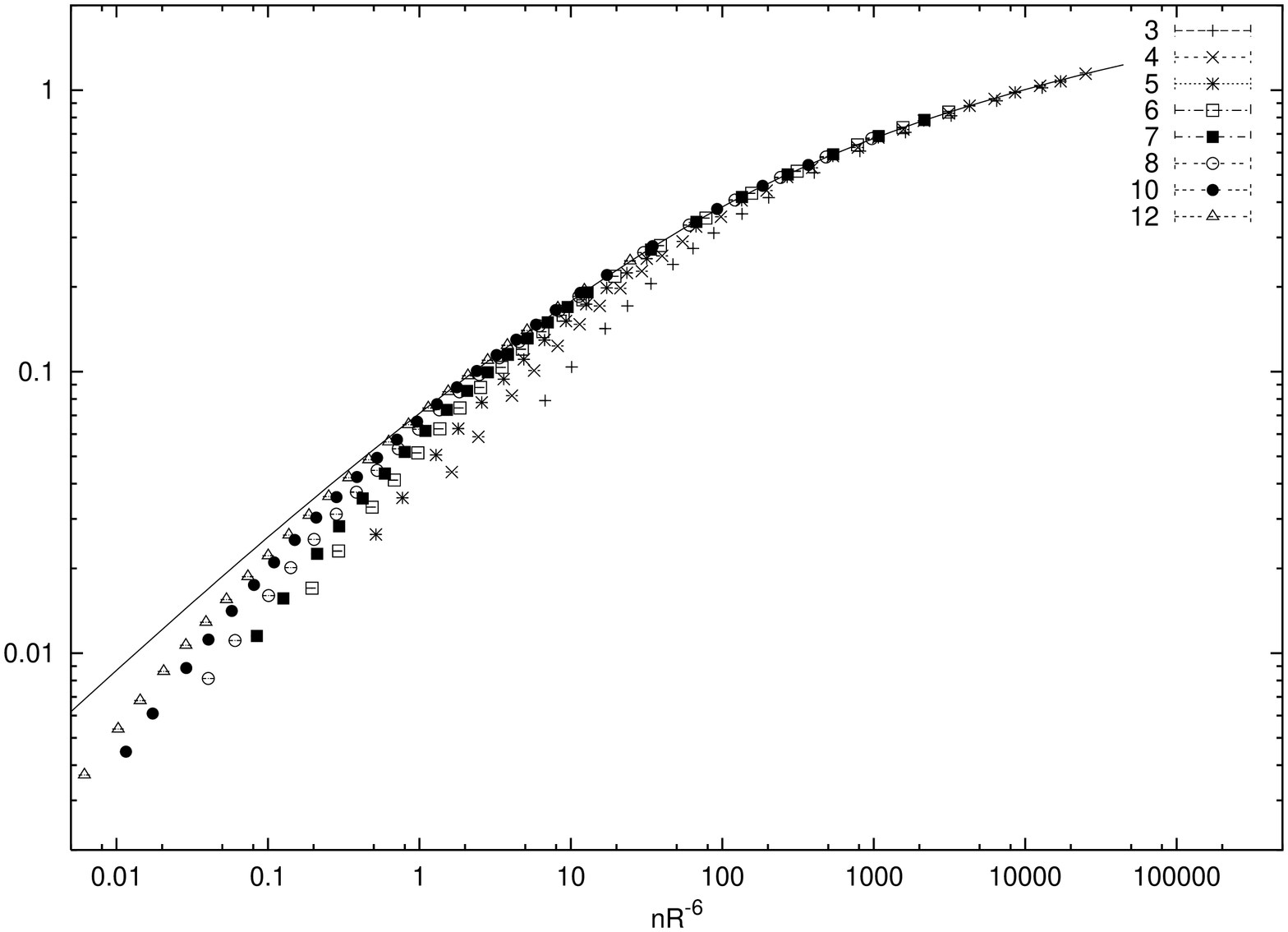}} \\
\centerline{\psfig{width=9truecm,angle=-90,file=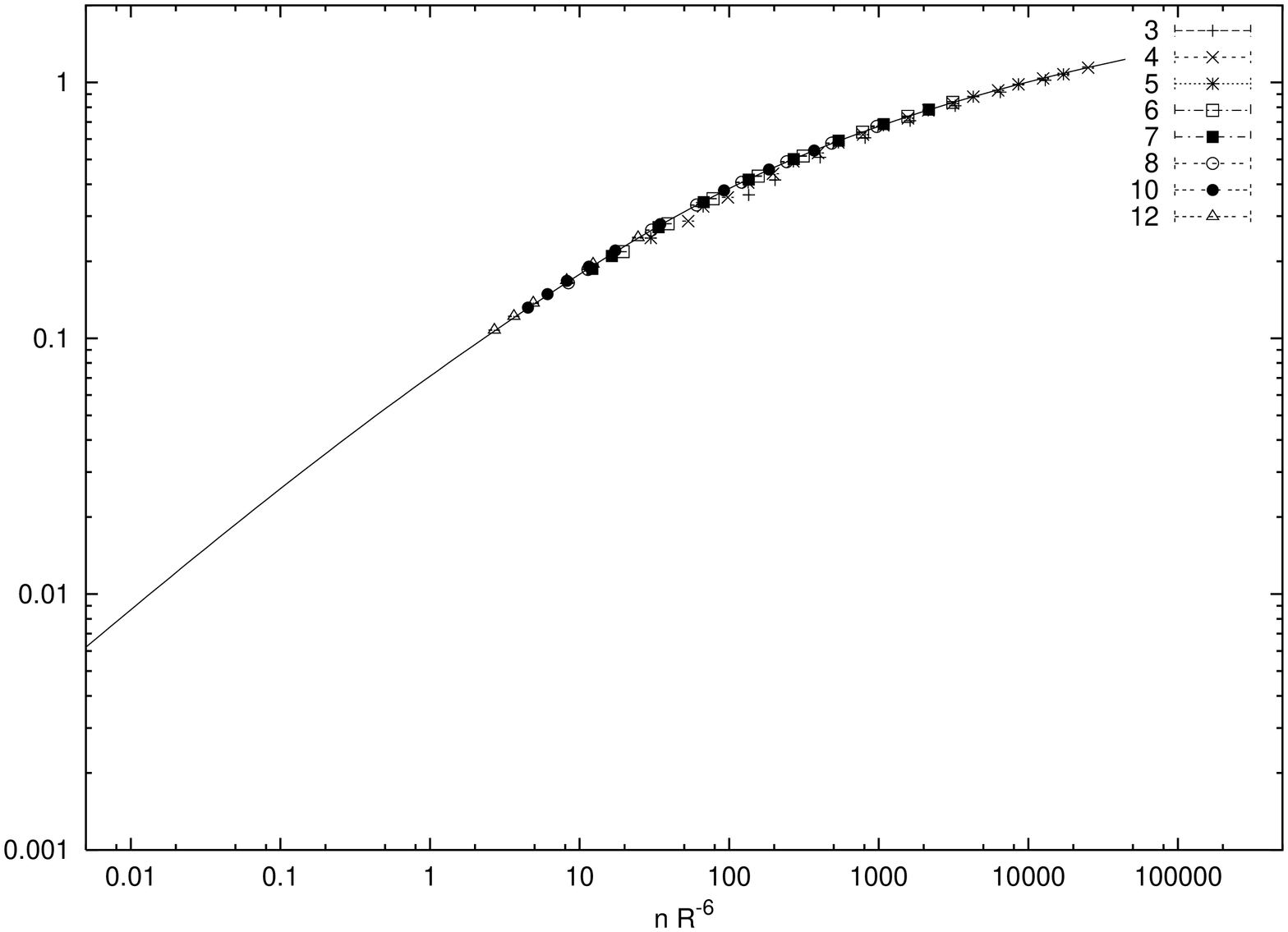}} 
\end{tabular}
\caption{Estimates of $\log \widetilde{c}_{n,R}$ versus $\widetilde{n}=nR^{-6}$.
The reported points correspond to $\rho=3,4,5,6,7,8,10,12$.
The line is the field-theoretical prediction. In the upper graph we report 
all points, in the lower one only those satisfying $n > V_R/2$.
}
\label{CNscal1}
\end{figure}

\begin{figure}
\centerline{\psfig{width=9truecm,angle=-90,file=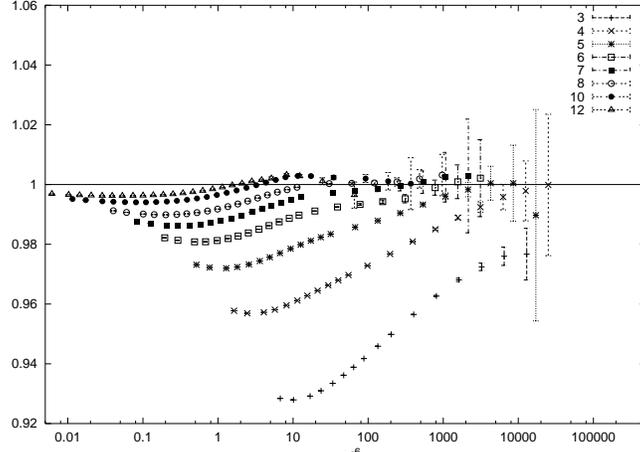}} 
\caption{Estimates of $\widetilde{c}_{n,R}/g_{c,\rm th}(\widetilde{n})$ 
versus $\widetilde{n}=nR^{-6}$.
The reported points correspond to $\rho=3,4,5,6,7,8,10,12$.
The errors only take into account the error on $\widetilde{c}_{n,R}$.
}
\label{CNscal2}
\end{figure}

\begin{figure}
\centerline{\psfig{width=9truecm,angle=-90,file=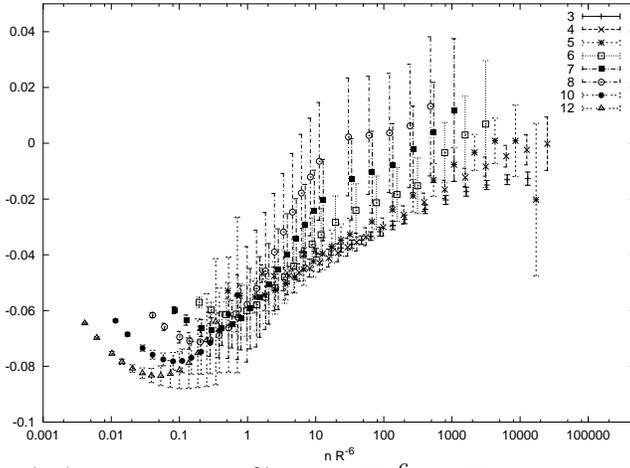}} 
\caption{Estimates of $\Delta_{c;n,R}$ 
versus $\widetilde{n}=nR^{-6}$.
The reported points correspond to $\rho=3,4,5,6,7,8,10,12$.
}
\label{CNscal3}
\end{figure}

\begin{figure}
\begin{tabular}{c}
\centerline{\psfig{width=9truecm,angle=-90,file=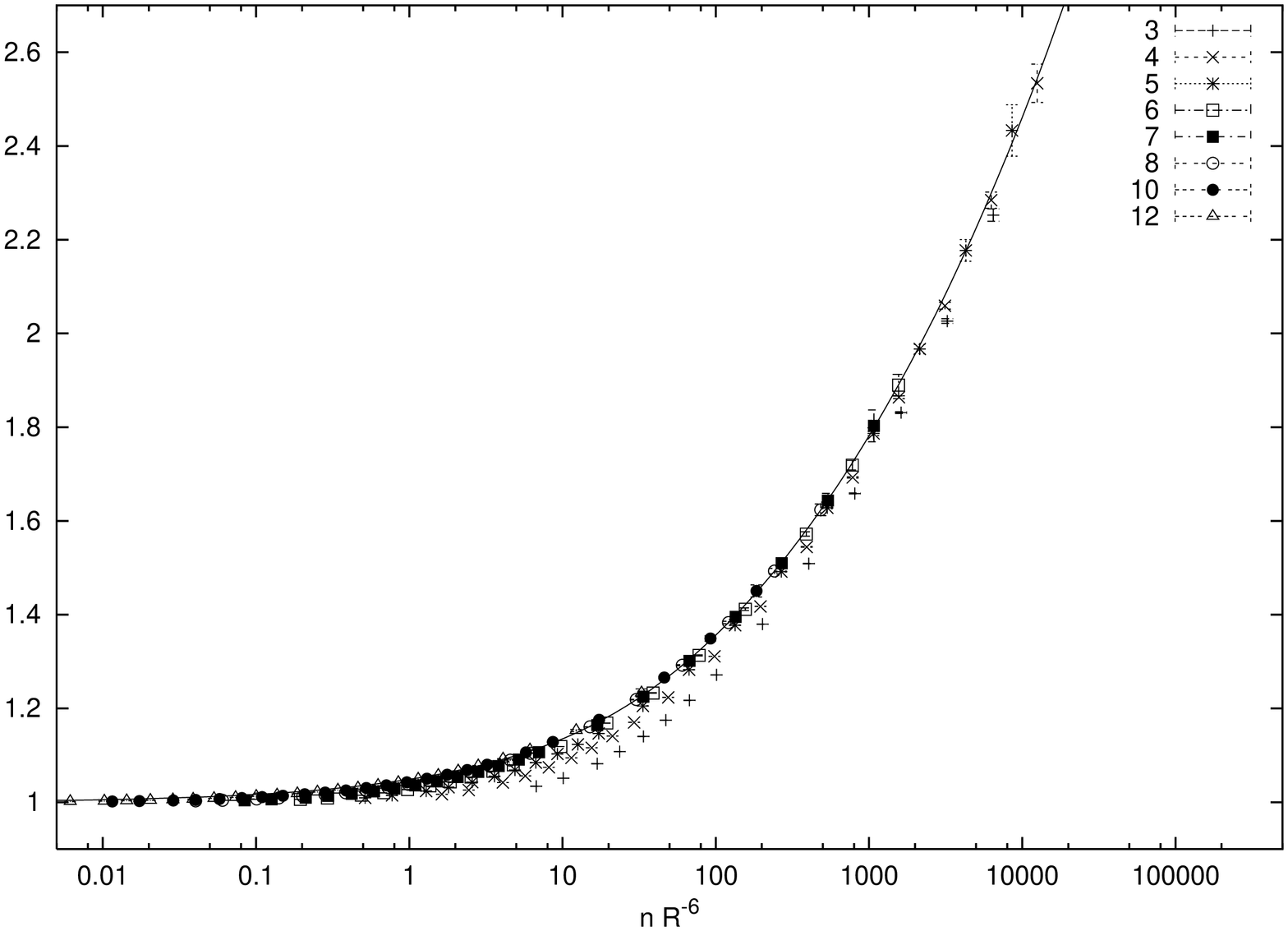}} \\
\centerline{\psfig{width=9truecm,angle=-90,file=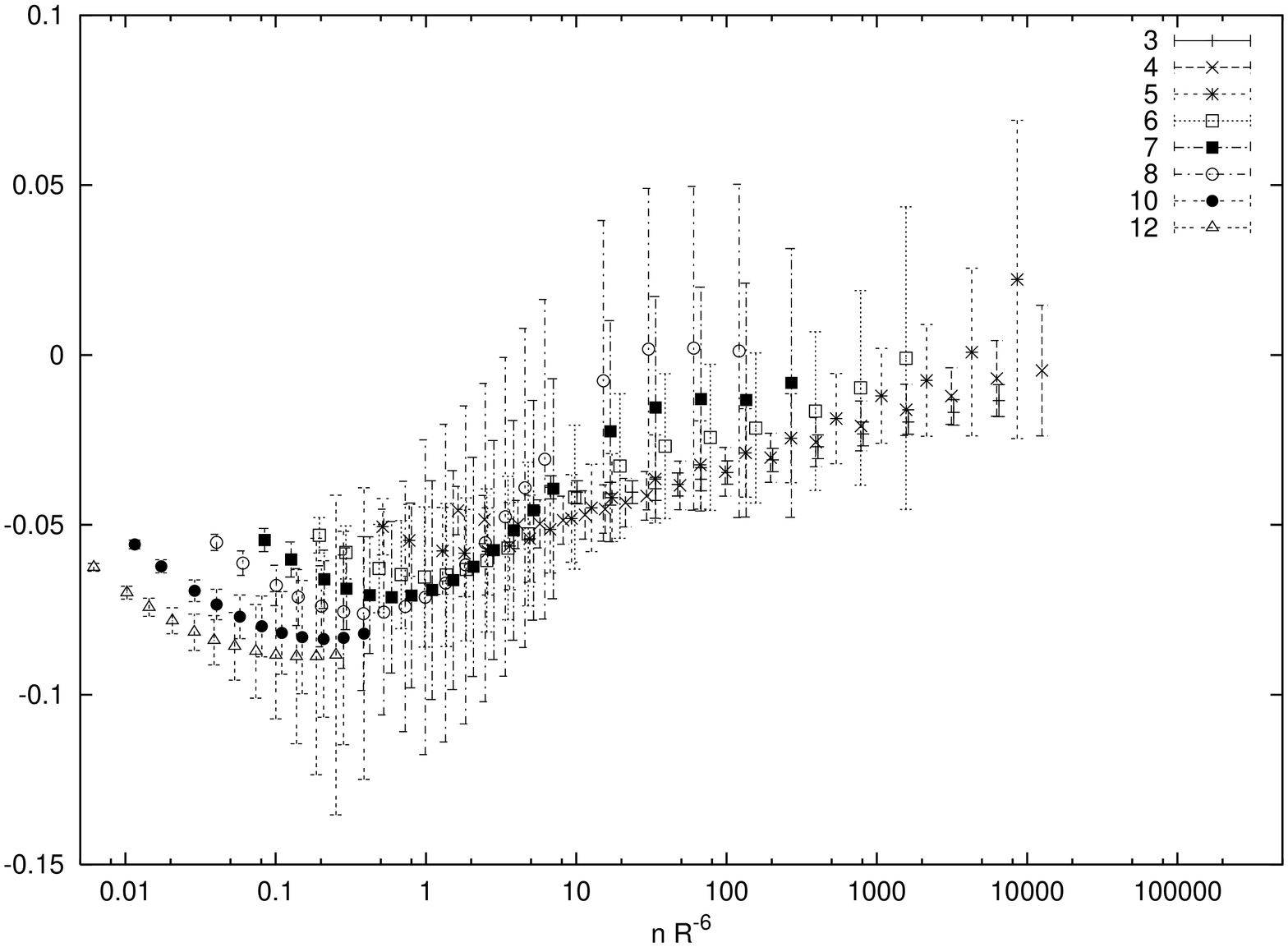}} 
\end{tabular}
\caption{Estimates of $Q_{n,R}$ (upper graph) 
and of $\Delta_{Q;n,R}$ (lower graph) versus $\widetilde{n}=nR^{-6}$.
The reported points correspond to $\rho=3,4,5,6,7,8,10,12$.
The continuous line in the upper graph is the theoretical prediction.
}
\label{Qscal}
\end{figure}

\begin{figure}
\centerline{\psfig{width=9truecm,angle=-90,file=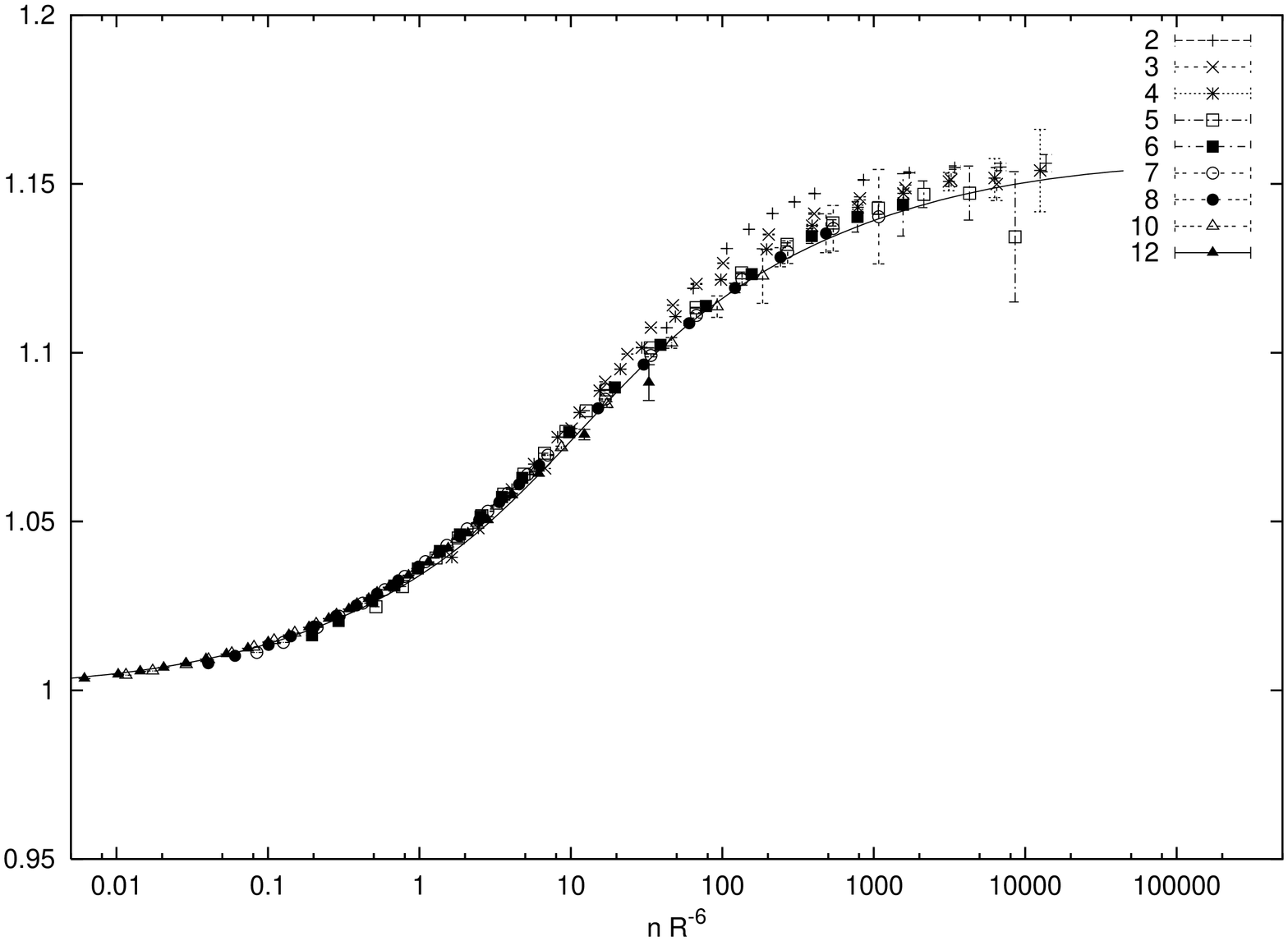}} 
\caption{Estimates of $\gamma_{\rm eff}({n},R)$
versus $\widetilde{n}=nR^{-6}$.
The reported points correspond to $\rho=2,3,4,5,6,7,8,10,12$.
The continuous line is the theoretical prediction.
}
\label{gammaeff}
\end{figure}

\begin{figure}
\begin{tabular}{c}
\centerline{\psfig{width=9truecm,angle=-90,file=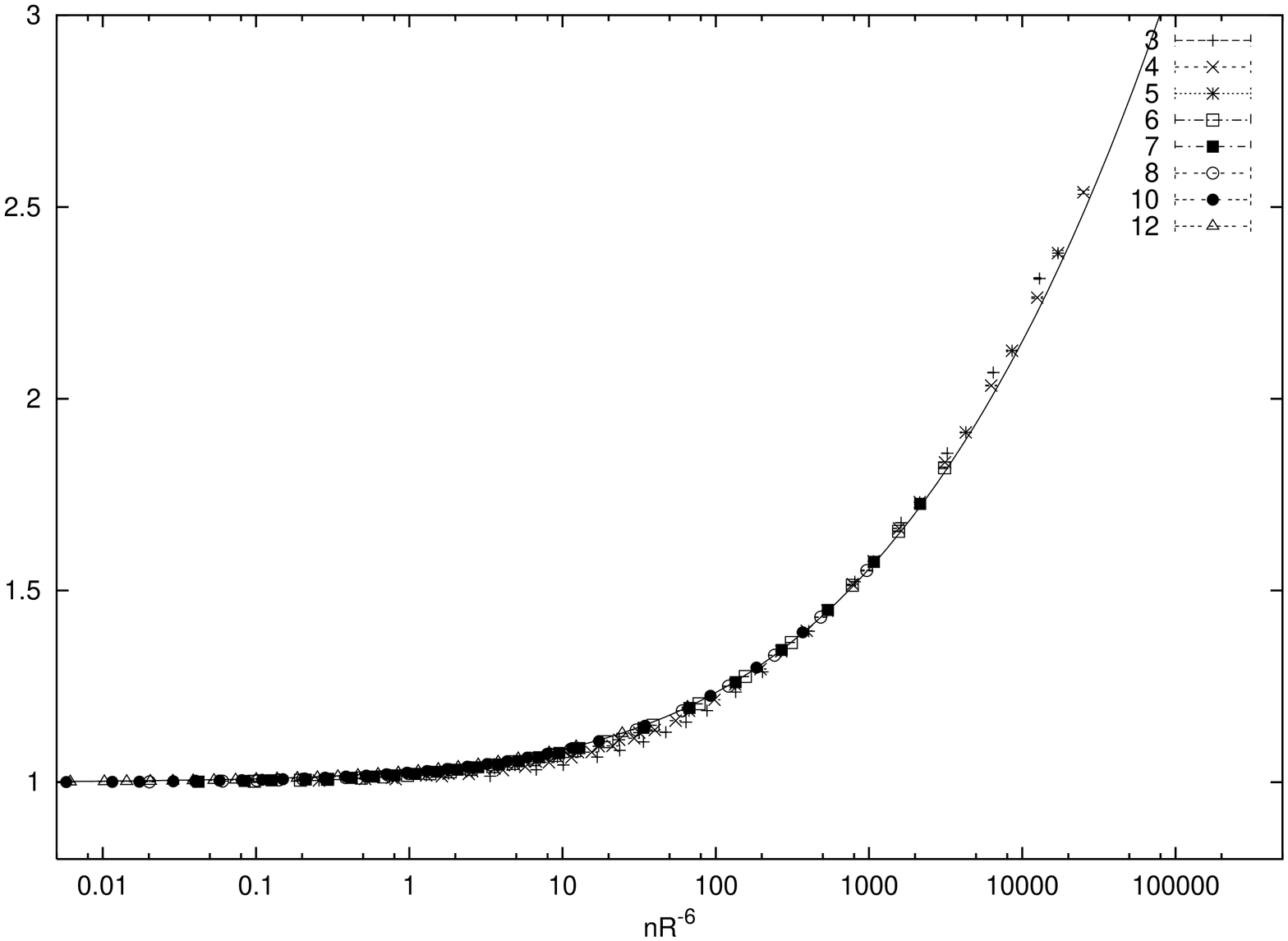}} \\
\centerline{\psfig{width=9truecm,angle=-90,file=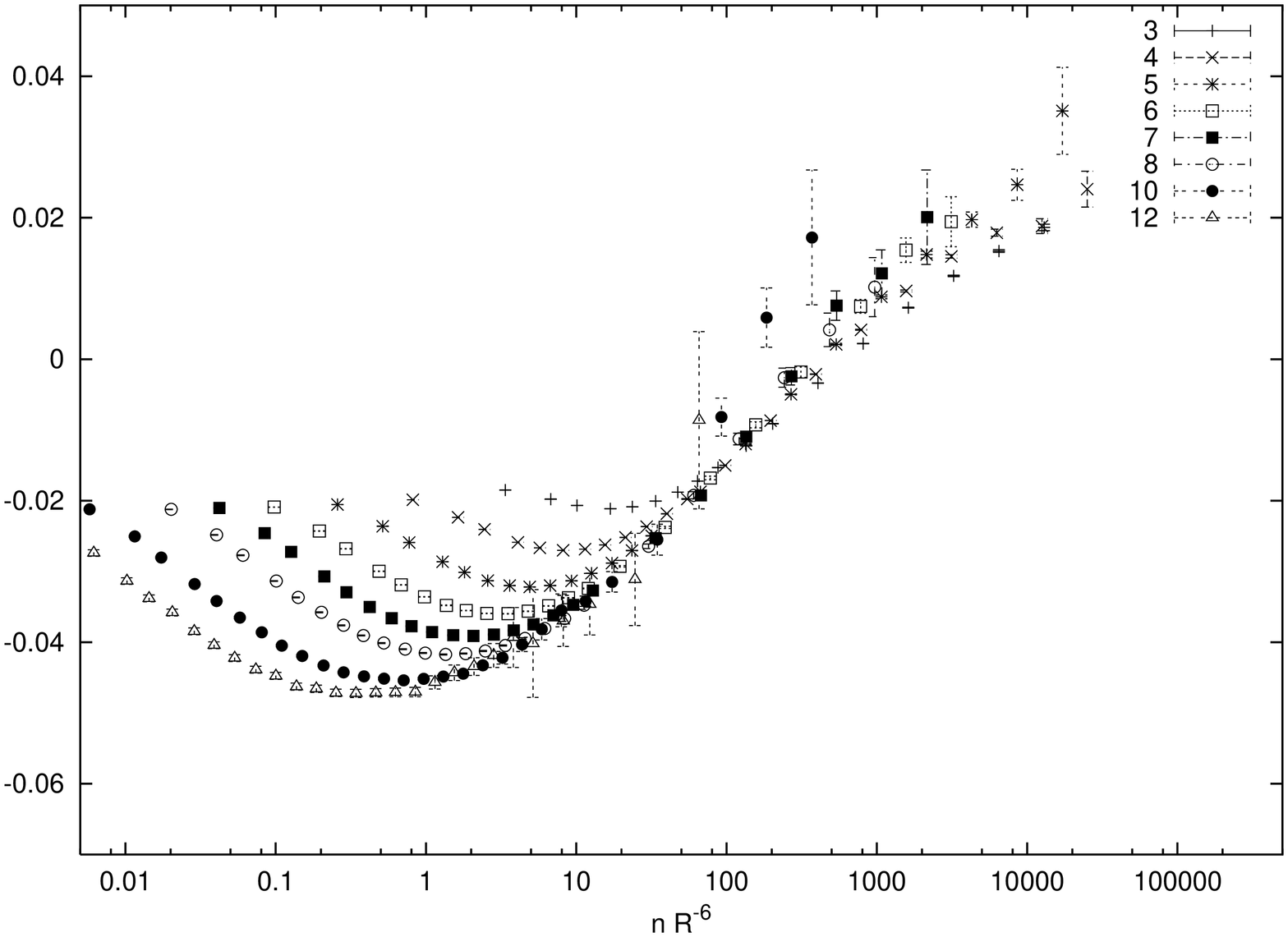}} 
\end{tabular}
\caption{Estimates of $\widetilde{E}^2_{n,R}/(6\widetilde{n})$ (upper graph) 
and of $\Delta_{E;n,R}$ (lower graph) versus $\widetilde{n}=nR^{-6}$.
The reported points correspond to $\rho=3,4,5,6,7,8,10,12$.
The continuous line in the upper graph is the theoretical prediction.
}
\label{Escal}
\end{figure}

\begin{figure}
\centerline{\psfig{width=9truecm,angle=-90,file=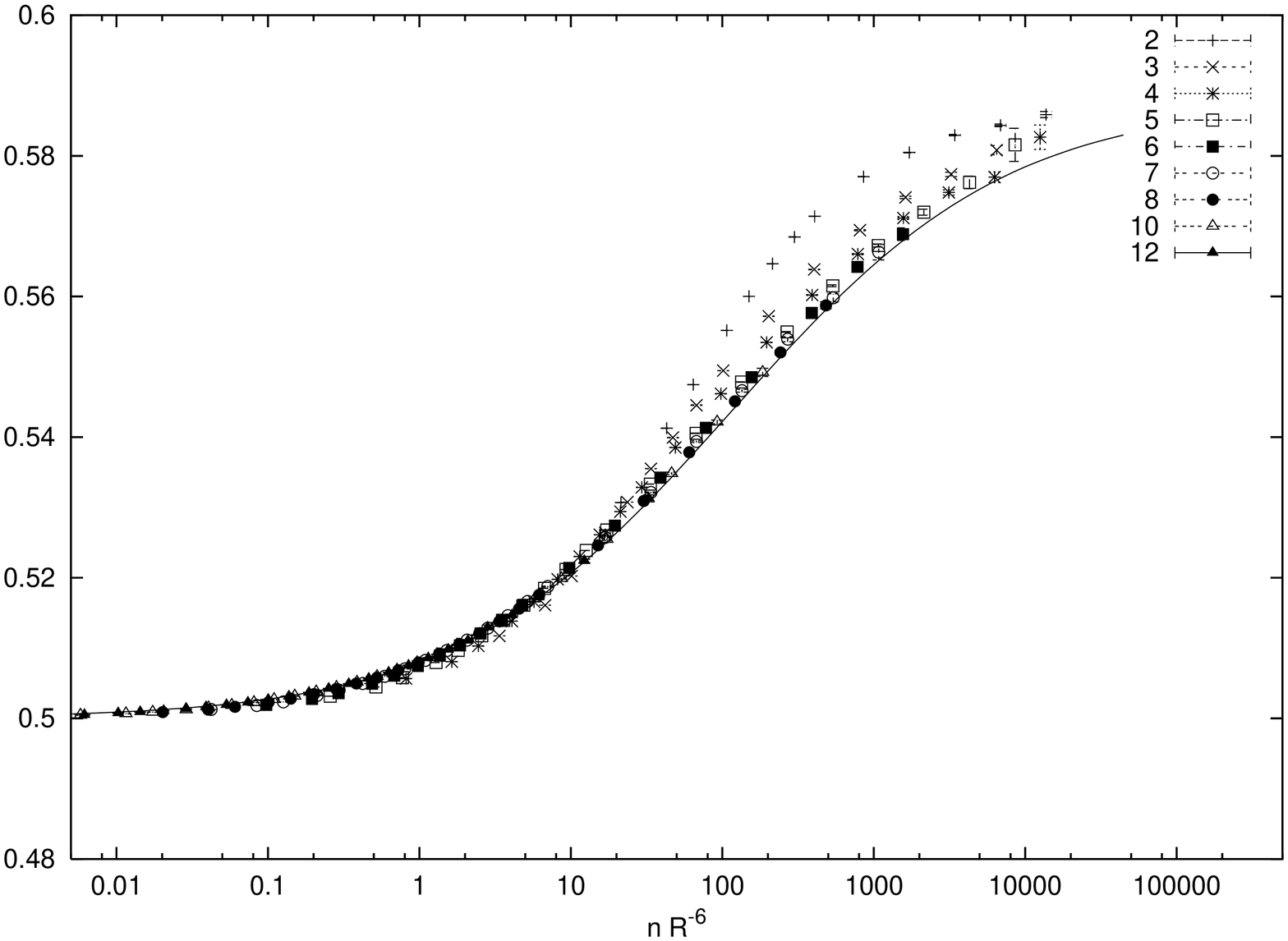}} 
\caption{Estimates of $\nu_{\rm eff}({n},R)$
versus $\widetilde{n}=nR^{-6}$.
The reported points correspond to $\rho=2,3,4,5,6,7,8,10,12$.
The continuous line is the theoretical prediction.
}
\label{nueff}
\end{figure}

\begin{figure}
\begin{tabular}{c}
\centerline{\psfig{width=9truecm,angle=-90,file=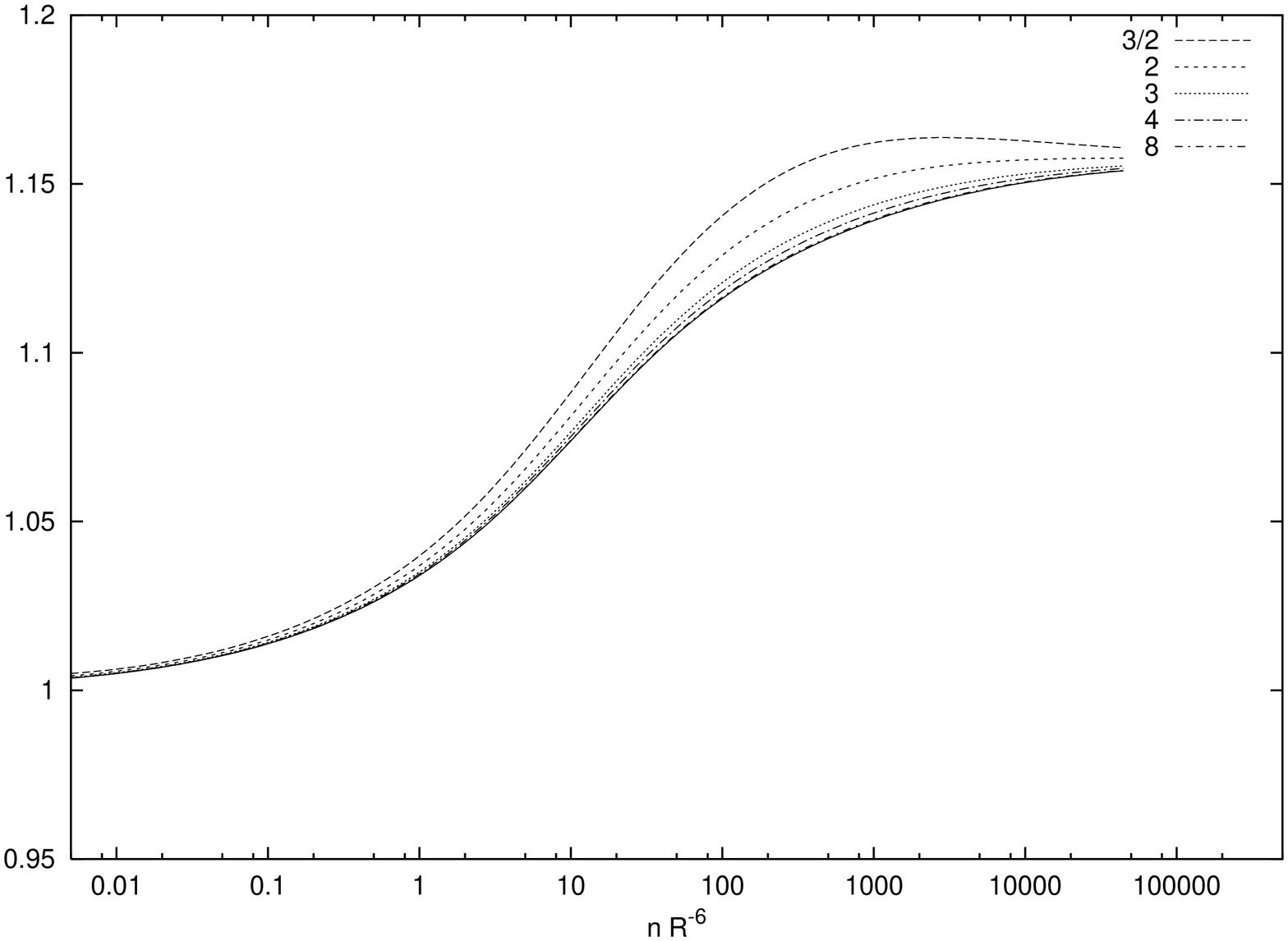}} \\
\centerline{\psfig{width=9truecm,angle=-90,file=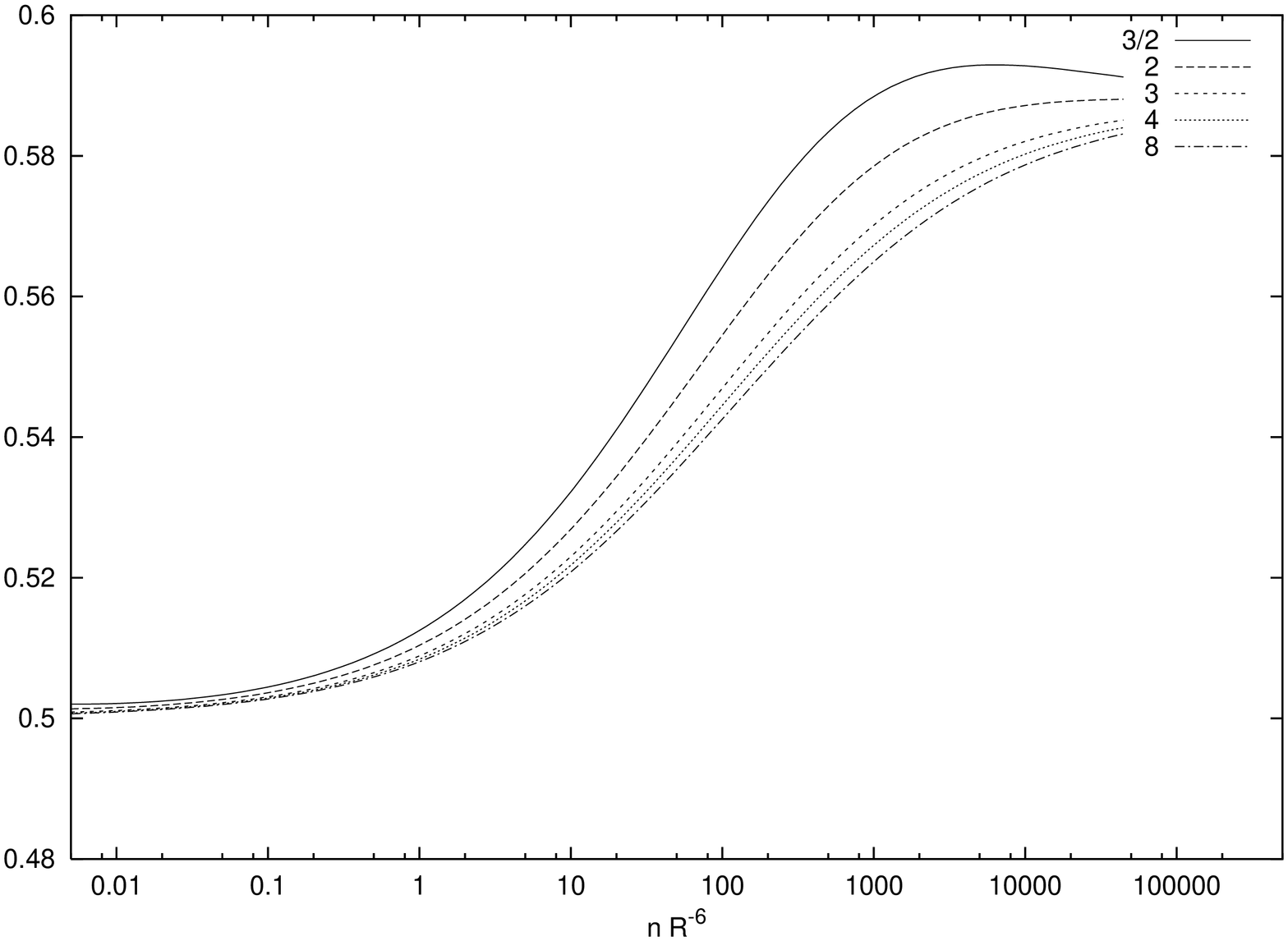}} 
\end{tabular}
\caption{The effective exponents $\gamma_{\rm eff}$ (upper curve) and 
$\nu_{\rm eff}$ (lower curve) $\widetilde{n}=nR^{-6}$, obtained from the phenomenological 
expressions \protect\reff{phenomenological1},  
\protect\reff{phenomenological2}, for several values of $\rho$.
}
\label{exp-phen}
\end{figure}

\end{document}